\documentclass[12pt,preprint]{aastex}

\newcommand{\be}{\begin{equation}}
\newcommand{\ee}{\end{equation}}

\newcommand{\msun}{M_\sun}
\newcommand{\msol}{M_\sun}
  
\newcommand{\simless}{\lower.5ex\hbox{$\; \buildrel < \over \sim\;$}}
\newcommand{\simgreat}{\lower.5ex\hbox{$\; \buildrel > \over \sim\;$}}
\def\lta{\,\raise 0.3 ex\hbox{$ < $}\kern -0.75 em
 \lower 0.7 ex\hbox{$\sim$}\,}
\def\gta{\,\raise 0.3 ex\hbox{$ > $}\kern -0.75 em
 \lower 0.7 ex\hbox{$\sim$}\,} 
\begin{document}

\title{A Statistical Stability Analysis of Earth-like Planetary Orbits in Binary Systems}

\author{Marco Fatuzzo,$^1$ Fred C. Adams,$^{2,3}$
Richard Gauvin$^1$ and Eva M. Proszkow$^{2}$} 
 
\affil{$^1$Physics Department, Xavier University, Cincinnati, OH 45207} 

\affil{$^2$Michigan Center for Theoretical Physics, University of Michigan \\
Physics Department, Ann Arbor, MI 48109}  

\affil{$^3$Astronomy Department, University of Michigan, Ann Arbor, MI 48109}

\email{fatuzzo@cerebro.cs.xu.edu, fca@umich.edu}

\begin{abstract} 

This paper explores the stability of an Earth-like planet orbiting a
solar-mass star in the presence of a stellar companion using $\sim
400,000$ numerical integrations.  Given the chaotic nature of the
systems being considered, we perform a statistical analysis of the
ensuing dynamics for $\sim 500$ orbital configurations defined by the
following set of orbital parameters: the companion mass $M_C$; the
companion eccentricity $e$; the companion periastron $p$; and the
planet's inclination angle $i$ relative to the stellar binary plane.
Specifically, we generate a large sample of survival times ($\tau_s$)
for each orbital configuration through the numerical integration of $N
\gg 1$ equivalent experiments (e.g., with the same orbital parameters
but randomly selected initial orbital phases). We then construct
distributions of survival time using the variable $\mu_s \equiv$ log
$\tau_s$ (where $\tau_s$ is in years) for each orbital configuration.
The primary objective of this work is twofold.  First, we use the mean
of the distributions to gain a better understanding of what orbital
configurations, while unstable, have sufficiently long survival times
to make them interesting to the study of planet habitability.  Second,
we calculate the width, skew, and kurtosis of each $\mu_s$
distribution and look for general features that may aid further
understanding and numerical exploration of these chaotic systems. To
leading order, most distributions are nearly Gaussian with a width
$\sigma \sim 0.5$, although the longest-lived systems display
substantial (non-Gaussian) tails. As a result, many independent
realizations of these systems must be considered in order to
characterize the survival time.  The situation is more complicated for
orbital configurations with longer mean survival times, owing in part
to the increasing importance of resonances. 

\end{abstract}



\section{INTRODUCTION}

Recent (and ongoing) discoveries of exoplanetary systems (e.g., Butler
et al. 1999; Marcy et al. 2001; Fisher et al. 2002; Tinney et
al. 2002; McArthur er al. 2004) have shown that Sun-like stars are
orbited by planets with a wide variety of orbital configurations.  Of
course, current planetary searches are biased toward large bodies with
short orbital periods, resulting in the discovery of planets with
masses ranging from $\sim 0.01 - 10 M_J$ and semi-major axis typically
ranging from $\sim 0.04 - 4$ AU.  It is expected that Earth-like
planets will also form alongside their Jovian counterparts (e.g.,
Ruden 1999; Lissauer 1993), even in binary stellar systems (e.g.,
Whitmire et al. 1998; Quintana et al. 2002; Lissauer et al. 2004).

The discovery of exoplanetary systems has spurred a renewed interest
in planetary dynamics, with a significant amount of attention being
given to the stability of specific multi-planet systems (e.g.,
Laughlin \& Adams 1999; S\"uli et al. 2005) and to the stability of
(hypothesized) terrestrial planets located within a specific system's
habitable zone (e.g., Noble et al. 2002; Asghari et al. 2004; Jones et
al. 2005; Haghighipour 2006).  In addition, the identification of
Jupiter-like planets in orbit around members of multiple star systems
(e.g., Eggenberger et al. 2004 and references therein) has also
provided an observational basis for studying planetary stability in
binary systems (e.g., Harrington 1977; Pendleton \& Black 1983; Rabl
\& Dvorak 1988; Holman \& Wiegert 1999; David et al. 2003;
Pilat-Lohinger et al. 2003; Musielak et al. 2005; Mudryk \& Wu 2006).
Of course, one must differentiate between stability within a habitable
zone and the more general study of overall system stability, with the
former imposing stricter conditions on a planet's allowed orbital
motion.

Planetary orbits in binary systems can be found with a wide range of
configurations.  The two most common orbital classes are P-type
orbits, in which the planet's semi-major axis is larger than that of
the binary so the planet orbits about both stars, and S-type orbits, in
which the planet revolves around one of the stars with a semi-major
axis smaller than that of the binary (e.g., Szebehely 1967, Dvorak et
al.  1989, Pilat-Lohinger \& Dvorak 2002). If the semi-major axis of
the planetary orbit is sufficiently large (for P-type orbits) or
sufficiently small (for S-type orbits), the orbital motion is stable
and well-ordered.  However, in binary configurations for which the 
stellar bodies come sufficiently close to the planet, the motion can
be chaotic and even unstable. One goal of this paper is to determine
the regime of binary parameter space for which S-type planetary orbits
are stable.  In the regime of parameter space near the border between
stability and instability, the orbits tend to be chaotic.

A complete analysis of planetary stability in stellar binaries must
reconcile the underlying chaotic nature of the systems being explored.
While theoretical conditions can be used to determine whether a
planetary system is unstable, it is still possible for such systems to
last for vast spans of time.  Numerical work must then be used to
delineate the boundaries of ``effectively'' stable parameter space.
However, for a given orbital configuration, the survival time $\tau_s$
of an unstable system (defined throughout this work as the number of
years it takes for the planet to either be ejected or collide with
either star) varies widely depending on the choices of initial orbital
phases. Because the systems are chaotic, this variation is not smooth,
i.e., small differences in the starting phase angles can lead to large
differences in the resulting survival times.  As a result, the
survival time for any set of binary properties can only be fully
characterized in terms of a distribution of output measures. Further,
as shown herein, the distribution of survival times has a substantial
width.  Even though the systems are chaotic in the regime of
interest, and hence display a wide distribution of survival times for
effectively equivalent starting conditions, we stress that the
distributions themselves are well-defined. As a result, the answer to
the question -- How long does a planet survive in any given binary
system? -- is a full distribution.  One goal of this work is to find 
such distributions for binary systems with intermediate 
semimajor axes $a \approx 1 - 50$ AU.

A statistical analysis of the stability of an Earth-like planet
orbiting a one solar-mass star in the presence of an outer-lying
companion was recently performed by David et al. (2003; hereafter
D03), leading to an estimate of the fraction ${\cal F}_b$ of binary
star systems that allow Earth-like planets to remain in the system
over a time scale of 4.6 Gyr.  Specifically, D03 found that ${\cal F}_b
\approx 0.5$ by calculating the survival time of an Earth-like planet
(with an initial circular orbit of R = 1 AU) for a range of companion
masses $M_C = 0.001 - 0.5 M_\sun$, initial eccentricities $e$, and
semi-major axis $a$.  For the parameter space explored by D03, the
planet's survival time (for a specified companion mass) depends most
sensitively on the companion's initial periastron distance $p =
a(1-e)$, but spans over two decades when sampled over a range of
semi-major axis values (with $e$ then set to give the desired value of
$p$).  Nevertheless, the mean value of log $\tau_s$ exhibits a clear
exponential dependence on the periastron.  It is important to note
that the range of survival times versus periastron presented in D03
results from both the chaotic nature of the three body problem and the
sampling over different ($a, e$) pairs for a given periastron value
(see \S 5.2 for a more complete discussion).
 
This paper extends previous work on the stability of an Earth-like
planet orbiting a Sun-like star in the presence of a stellar
companion.  Although much of the previous work focused on coplanar
systems (e.g., D03), this work explores the full range of inclination
angles.  Specifically, we explore $\sim 500$ different orbital
configurations (defined through the choice of the companion mass
$M_C$, eccentricity $e$, and periastron $p$, and the planet's
inclination angle $i$ relative to the stellar orbital plane), and we
broadly sample the orbital parameter space inhabited by most observed
binary systems. Our main body of work explores $376$ different orbital
configurations organized in four series (1 -- 4). An additional 136
orbital configurations with low inclinations, organized in three
series (5 -- 7), are performed to further explore some of the detailed
structure exhibited by the output measures of the first four series.
We generate a large sample of survival times ($\tau_s$) for each
orbital configuration through the numerical integration of $N \gg 1$
equivalent experiments (with the same basic orbital parameters but
randomly selected initial orbital phases), performing a total of $\sim
390,000$ separate orbital simulations; this sample of orbital
simulations is thus an order of magnitude larger than our previous
study (D03). This broad survey of parameter space provides a cleaner
delineation of the orbital configurations that (while potentially
unstable) have sufficiently long survival times to be interesting for
planetary habitability (as noted above, however, stability within a
planet's habitable zone imposes additional constraints on the planet's
orbital motion).

Another focus of this new work is the characterization of the
distributions of survival times.  As discussed earlier, the regime of
parameter space near the stability/instability border is chaotic and
the results must be described statistically.  Performing multiple
realizations for each orbital configuration is a necessity.  This
paper constructs distributions of survival times --- using the
variable $\mu_s \equiv$ log $\tau_s$, with $\tau_s$ defined in years
--- for each orbital configuration under consideration. The resulting
distributions are then characterized by their mean, width, skew and
kurtosis.  We also calculate the fraction of runs that lead to the
planet's ejection from the system (and the fraction accreted by one of
the stars). With the construction of these distributions, we can study
how they vary over the regime of binary parameter space; these
distributions also provide important guidance for future work (e.g.,
the width of the distribution determines how many independent
realizations of the numerical problem are necessary to achieve a
desired accuracy in estimating the mean value).

The paper is outlined as follows: We discuss the numerical method used
in our work in \S 2.  We present the results of our numerical work for
series 1 -- 4 in \S 3, and characterize the resulting distributions
for these series in \S 4.  We present the results of series 5 -- 7 in
\S 5, and use the results of these runs to explore certain aspects of
the rich structure exhibited by these three body-systems.
Specifically, we: 1) consider the effect of integer ratios of initial
planet/companion orbital periods -- a necessary but not sufficient
condition for resonance (see, e.g., Murray \& Dermott 2000); 2)
explore more fully the dependence of ejection time on eccentricity and
on periastron; 3) characterize the fraction of ejection events; and 4)
consider the stability exhibited by high inclination, low eccentricity
orbits.  We present our conclusions in \S 6.

\section{NUMERICAL SCHEME}  

We present here the numerical method by which we calculate the
survival time of an Earth-like planet orbiting a Sun-like star in the
presence of a stellar outer companion.  Through long term dynamical
interactions with the outer companion, the orbital elements of the
Earth-like planet evolve, generally in chaotic fashion, until the
planet is either ejected from the system or collides with one of the
two stars.  In order to explore this stability issue on intermediate
time scales, Newton's equations of motion are integrated directly
using a Bulirsh-Stoer (B-S) scheme (Press et al. 1992).  Although
direct integration is computationally more expensive (e.g., compared
to symplectic integration), it is accurate and explicit. For the
systems at hand, our B-S scheme incurs errors in relative accuracy of
order 1 part in $10^{11}$ per total time step, where each time step in
the three-body problem is variable, but has a typical value of about
10 days.  The accumulated error for a given integration is
characterized through the ratio of $\Delta E / E$, where $E$ is the
initial system energy (which should be conserved) and $\Delta E$ is
the difference between this value and the final energy of the
numerically integrated system.\footnote{The error accumulation
involves a random walk process so that high accuracy can (usually) be
maintained even when the product of the error per time step and the
number of time steps exceeds unity.}  Since the ratio of the
Sun-planet to Sun-companion orbital energy is $\sim (M_p/M_C) (a_C /
a_p) \sim 10^{-5}$ for the systems that yield the longest survival
times in our study (where $a_p$ is the Earth's semi-major axis), we
check {\it a posteriori} that $\Delta E / E$ remains less than $\sim
10^{-6}$ to ensure that accumulated errors do not affect our results.

The planet's initial orbit is always set to be circular ($e_p$ = 0)
with radius $R=a_p=1$ AU. The companion mass $M_C$, eccentricity $e$,
and periastron $p$, as well as the planet's initial inclination angle
$i$ (relative to the stellar binary plane) are then specified for each
run, and the system is integrated forward in time. For the sake of
definiteness, and in order to cover a large range of parameter space,
we use an upper limit integration time of $\tau_{run} = 10^8$ years
for the runs presented in \S 3 (series 1 -- 4), and $\tau_{run} =
10^9$ years for the runs presented in \S 5 (series 5 -- 7).  Our
experiments therefore give us either a time scale for survival or a
lower limit of $\tau_{run}$ on the possible survival time.  The planet
is considered to be ejected if any of the following conditions are
met: The energy of the planet becomes positive; the eccentricity of
the planet exceeds unity; or the semi-major axis of the planet exceeds
a maximum value (taken here to be 100 AU).  The planet is considered
to collide with the solar mass star if the periastron of the planet
becomes smaller than the stellar radius (assumed to be 1 $R_\odot$) so
the planet is accreted, and to collide with the companion if its orbit
crosses within one radius of the companion's center-of-mass position
(although for the orbital configurations considered here, the latter
result is effectively ruled out).  The survival time for a given
orbital configuration is explored statistically by sampling over $N$
equivalent realizations set through the random assignment of the
initial orbital phase angles, where $N = 10^3$ for series 1 -- 4
(presented in \S 3) and $N = 10^2$ for series 5 -- 7 (presented in \S
5).

\section{RESULTS OF PARAMETER SPACE SURVEY}  

This section presents the results of $376,000$ numerical experiments
with orbital parameters organized into four series with the following
[$M_C$, $e$] values: (1) [$0.1 M_\sun$, 0.5]; (2) [$0.5 M_\sun$, 0.5];
(3) [$0.5 M_\sun$, 0.75]; and (4) [$0.5 M_\sun$, 0.25].  Values for
the companion periastron and the planet's inclination angle relative
to the Sun-companion orbital plane range from $p = 2 - 6$ AU and $i =
0^o - 90^o$, respectively, for each series.\footnote{Although
high-inclination configurations have been included for completeness,
observations suggest that planetary systems are likely to form in
binaries with relatively small ($i \simless 20^o$) inclination angles.
For example, disks in binary systems are close to coplanar (e.g.,
Mathieu et al. 1991).}  A total of $N = 10^3$ equivalent realizations
(with initial phase angles sampled randomly) were performed for each
of the 376 different orbital configurations explored (defined by the
parameter space four-vector [$M_C, e, p, i$]).
  
For the chaotic systems being explored herein, different choices of
the initial phase angles can lead to widely different dynamical
behavior, and hence the values of survival time $\tau_s$ can differ by
orders of magnitude for effectively equivalent starting states.  The
survival time for a given orbital configuration is best characterized
in terms of a distribution in $\mu_s \equiv $ log $\tau_s$ (where the
survival time is expressed in years).  Each orbital configuration is
therefore characterized by the mean or expectation value $\langle
\mu_s \rangle$, the width
\be
\sigma = \left[{\Sigma (\mu_s - \langle\mu_s\rangle)^2 \over N}\right]^{1/2}\;,
\ee
the skew
\be
\hbox{\rm sk} = \left[{\Sigma (\mu_s - \langle\mu_s\rangle)^3 \over N \sigma^3}\right]\;,
\ee
and the kurtosis
\be
\hbox{\rm ku} = \left[{\Sigma (\mu_s - \langle\mu_s\rangle)^4 \over N \sigma^4}\right] - 3 \;,
\ee
of its corresponding distribution.  A secondary output measure also
explored here is the fraction $f_e$ of ejection events, defined as the
number of runs that result in the planet being ejected from the system
divided by the total number of ejection/accretion events (see \S 5 for
further discussion).

The calculated values of $\langle\mu_s\rangle$ are plotted versus
inclination angle $i$ for several periastron tracks in Figure 1, with
series 1 -- 4 separated into panels (a) -- (d).  For all cases, a
clear transition between longer and shorter lived systems occurs at $i
\sim 40^o$, in agreement with results of previous studies (e.g.,
Harrington 1977; Pendleton \& Black 1983; Innanen et al. 1997;
Haghighipour 2006).  Additionally, a clear region of stability is
found at $i \approx 60^o$ for series 4.  Note that Kozai resonances
occur for large inclination angles (see \S5 for further analysis and
discussion).

Clear trends are exhibited by our results.  As expected, survival time
depends strongly on the companion mass and periastron, and somewhat
weakly on inclination angle for $i \simless 40^o$.  In addition,
panels (b) -- (d) in Figure 1 corresponding to series 2 -- 4 ($M_C =
0.5 \msun$; $e = 0.5, 0.75$ and $0.25$) indicate that while survival
times increase with eccentricity for periastron values below $p \sim
3$ AU (as one would naively expect given the longer orbital period of
the companion orbits), the opposite trend is suggested for periastron
values greater than 3 AU.  The dependence on survival time on
eccentricity will be further explored in \S5.  The robust structure of
orbital systems (being chaotic in nature) is also clearly illustrated
by the interweaving of the $p = 2.5$ AU and $p = 3$ AU tracks in panel
(a) of Figure 1.  Naively, one expects the survival time to increase
with increasing periastron, and this general trend is clearly seen in
all four panels of Figure 1. The enigmatic behavior of the $p = 3$ AU
case in series 1 thus suggests some type of resonance behavior.
Indeed, this set of orbital parameters corresponds to an initial
companion orbital period exactly 14 times larger than the planet's, a
necessary but not sufficient condition for resonance to occur (e.g.,
Murray \& Dermott 2000).  We more fully explore the structure
associated with integer ratios of initial orbital periods in \S 5.

One goal of this work is to determine what region of orbital parameter
space may be of interest to the study of planet habitability.  It is
well known that many planetary systems, while Hill unstable, can
survive long enough to make them effectively stable on time scales
comparable to the age of the Universe (but as previously noted, the
stability criteria explored in this paper is not the same as stability
within a habitable zone).  Extrapolating the results of our work
provides an estimate of the minimum companion periastron values
$p_{min}$ required for an Earth-like planet orbiting a Sun-like star
to remain stable in a stellar binary system for $\sim$ 5 Gyrs (a time
scale comparable to the current age of the Solar System). Doing so in
the low inclination angle regime leads to an estimate of $p_{min} \sim
4$ AU for a companion mass of $M_C = 0.1 \msun$ (see also Figure 8 and 
\S 5), and $p_{min} \sim 6$ AU for $M_C = 0.5 \msun$ (see also Figure
13 and \S5).  These results are in good agreement with the conservative
estimate of $p \sim 6-7$ AU for a 4.6 Gyr survival time obtained by D03. 

\section{WIDTHS OF THE DISTRIBUTIONS}

The characterization of the survival times through the output measure
$\mu_s \equiv $ log $\tau_s$ was motivated by the large dynamic range
of survival times exhibited by different realizations of the same
orbital configuration.  Although the use of a logarithmic time unit is
arbitrary, it appears to be robust for the cases explored in series 1
-- 4.  Specifically, the distributions of $\mu_s$ for most of the
orbital configurations presented in \S 3 are nearly Gaussian to leading
order, i.e., $\tau_s$ is log-normally distributed; further, the
distributions have roughly the same widths $\sigma \sim 0.5$.  To
further illustrate this point, panels (a) -- (c) of Figure 2 show the
distributions of $\mu_s$ values obtained for orbital configurations
with $M_C = 0.1 M_\sun$, $p = 2.5$ AU, $e = 0.50$ and $i = 0^o$,
$25^o$, and $50^o$, along with corresponding Gaussian distributions
(solid line) with the same computed mean, width, and
normalization.\footnote{The degree to which a given distribution is
non-Gaussian is measured by its skew and kurtosis.  The skew reflects
how symmetric about the mean value a distribution is, with
``right-heavy'' distributions having positive skews.  The kurtosis
reflects how peaked or flattened a distribution is, with flatter than
normal distributions having a positive kurtosis, and pointier than
normal distributions having a negative kurtosis (e.g., Press et
al. 1992). }

Interestingly, distributions in our survey with $\sigma \simgreat 0.7$
also appear to have Gaussian peaks with widths $\sim 0.5$, where the
presence of tails leads to the higher computed values of the total
widths.  As an example, the distribution for the orbital configuration
$M_C = 0.5 M_\sun$, $p = 3.5$ AU, $e = 0.50$ and $i = 35^o$ is shown
in Figure 2, panel (d).  Again, the solid line represents a Gaussian
profile with the same width, mean and normalization as the computed
distribution.  Clearly, the presence of a tail above the peak makes
the overall profile wider and non-Gaussian.

One of the most important aspects of this work is the characterization
of the $\mu_s$ distribution widths.  These values quantify the degree
of chaos in these three-body systems and allow one to determine the
precision of the mean value $\langle\mu_s\rangle$ calculated from $N$
equivalent numerical experiments.  For example, if the parent
distribution is nearly Gaussian, then the mean value calculated from
$N$ equivalent numerical experiments has a 68\%, 95\% and 99\% 
probability of being within 1, 2, and 3 $\sigma / \sqrt{N}$,
respectively, of the true mean.

The dependence of the distribution width on inclination angle and on
mean survival time (characterized by $\langle\mu_s\rangle$) for the
distributions of the 376 orbital configurations in series 1 -- 4 is
illustrated by the scatter plots presented in panels (a) and (b) of
Figure 3.  The solid line in panel (a) connects the mean widths
calculated at each inclination angle, determined by combining all of
the results from orbital configurations with the same inclination
angle (but offset by the mean so as to center each distribution about
zero) to form a single distribution, and then calculating the ensuing
width as per equation (1).  Most of the distributions for the orbital
configurations considered in this section have widths in the range
$\sigma \sim 0.2 - 0.6$.  In addition, the distribution width has only
a weak dependence on inclination angle, and no clear trend that
differentiates the four series from each other.  The distribution
width also seems independent of the mean survival time for orbital
configurations with $\langle \mu_s\rangle \simless 4$, although wider
distributions appear increasingly likely when $\langle \mu_s \rangle
\simgreat 4$.  This result is further explored in \S 5.

A proper interpretation of the meaning of a width calculated from a
sample distribution requires an underlying assumption about the parent
distribution from which the sample was drawn.  To assess the meaning
of the values of $\sigma$ calculated for the orbital configurations in
series 1 -- 4, we need to gauge whether a Gaussian approximation for
the calculated $\mu_s$ distributions is warranted.  Toward this end,
we present scatter plots of skew and kurtosis versus width in panels
(c) and (d) of Figure 3 for the distributions of the 376 orbital
configurations of series 1 -- 4.  The dotted lines in these figures
represent the three-sigma values (for the parameter relevant to the
given panel) calculated through random sampling (with $N = 10^3$) of a
Gaussian parent distribution. For example, the skew of a sample
distribution built-up by randomly sampling a Gaussian parent
distribution $10^3$ times would have a 99\% probability of falling
within the dotted lines of Figure 3, panel (c).  Clearly, the
overwhelming number of orbital configurations lead to distributions
with positive skew and kurtosis, with most skew values ranging between
0 and 2, and most kurtosis values ranging from 0 and 5. No obvious
trends differentiate series 1 -- 4 in this respect. As a point of
reference, we note that the distributions shown in Figure 5 (plotted
in terms of ln $P$) have values of skew between 1 and 1.2, and values
of kurtosis between 5 and 8.  Clearly, while the distributions
explored in this section are not perfectly Gaussian, they nevertheless
have Gaussian-like peaks, and to first order, can be reasonable well
approximated by Gaussian profiles. Furthermore, the departure of the
distributions from a Gaussian form occurs through a tail at high
values of $\mu_s$.
 
To gain further insight into the distribution widths expected from
this class of systems, we produce histograms of survival times for
subsets of the orbital configurations in series 1 -- 4, with each
survival time normalized about the mean of its own distribution (i.e.,
all of the distributions are normalized so that their mean is zero,
and then combined to form new distributions for specific subsets of
the data).  Figure 4 shows the resulting histograms for each series,
plotted in terms of the natural log of the probability $P$ and
normalized values of $\mu_s$. Figure 5 presents the resulting
histograms for series 2 -- 4 (panel [a]) and for various angle subsets
of series 2 -- 4 (panels [b] -- [d]).  The width of each distribution
is included in the figures; the skew and kurtosis for each of the
histograms range from 1 to 1.6, and from 3 to 8, respectively.  In all
cases, power exists in the wings, but a clear peak is always observed.
Figure 6 presents probability histograms of $\sigma$ for all of the
distributions within each of the four series (these histograms thus
represent the distribution of distribution parameters).  The
calculated mean width $\langle\sigma\rangle$ and distribution width
$\sigma_\sigma$ are denoted in each panel.  We note that the
calculated values of $\langle\sigma\rangle\,\sim 0.35$ are slightly
lower than the mean of the distributions shown in Figures 4 and 5.
Similar histograms for all of the distributions in series 2 -- 4 as
well as the low angle ($i \le 30^o$) subset are presented in Figure 7.

For the orbital configurations explored in this section, distributions
have typical widths in the range $\sigma \sim 0.2 - 0.6$.  Taken at
face value, a ``typical'' distribution of $\mu_s$ for the most likely
orbital configurations found in our galaxy (i.e., $i \simless 20^o$)
is expected to have a width of $\sigma \sim 0.5$ (see the discussion
of \S 5.3 for further complications).  As a result, the value of
$\langle\mu_s\rangle$ calculated through a random sampling with a
sample size of $N$ would have a 68\% probability of being within $\sim
0.05\, \sqrt{(100/N)}$ of the true value, and a 99 \% probability of
being within $\sim 0.15\, \sqrt{(100/N)}$ of the true value, assuming
that the parent distribution is nearly Gaussian.  This result provides
important guidance for future studies on the survival time of
planetary systems.

\section{STRUCTURE IN THE OUTPUT VARIABLES}

The chaotic nature of three-body problems can lead to a rich amount of
structure in the output variables that characterize the underlying
dynamics.  Indeed, a fair amount of structure is evident in the output
measures presented in \S3.  We explore certain aspects of that
structure here through three additional series of runs.  Specifically,
we consider: 1) the structure resulting from integer ratios of the
initial companion/planetary orbital periods; 2) the dependence of
survival time on periastron and eccentricity; 3) structure in the
output measure $f_e$ -- the fraction of simulations for a given
orbital configuration that lead to the planet's ejection, and 4) the
structure at $\sim 60^o$ evident in panel (d) of Figure 1.  The
numerical experiments presented in this section (series 5 -- 7) were
integrated to a maximum run time of $\tau_{run} = 10^9$ years.  To
compensate for this increase, the number of equivalent experiments
performed for each orbital configuration was reduced to $N = 10^2$.
Based on the results of \S 4, we still expect good statistical
results.  The overall goal of this section is to explore further the
aforementioned observed structure in the output variables, to provide
some insight into the underlying mechanisms, and to relate this work
to results of previous studies of orbital dynamics.  


\subsection{Mean motion orbital resonances}

As noted in \S3, the mean value of $\mu_s \equiv $ log $\tau_s$
generally increases with increasing periastron.  This result is
expected, given that an increase in periastron corresponds to both an
increase in distance of closest approach between the planet and
companion and an increase in the companion's orbital period $P_b$ (for
a constant eccentricity).  However, orbital configurations in series 1
with $p = 3.0$ AU and inclination angles $25^o \le i \le 60^o$ have
$\mu_s$ distributions with smaller means than their $p = 2.5$ AU
counterparts, and distributions with smaller means than their $p =
2.9$ AU counterparts at all but the largest inclination angles.  As
noted in \S3, this result suggests the presence of a resonance
condition at $p$ = 3.0 AU, and indeed, the companion and planet's
orbital periods are in a ratio of $n_C:n_p = 14:1$.

The presence of resonance conditions can play an important role in the
dynamics of a planetary system -- and certainly do so in our own solar
system (see, e.g., Murray \& Dermott 2000 for a thorough discussion).
While an extensive review of resonances is beyond the scope of this
paper, we note that an integer ratio of initial orbital periods of the
planet and companion star can have either a stabilizing or
destabilizing effect on the system, depending on the initial positions
of the two objects.  For example, for a $2:1$ orbital configuration,
if the planet and the companion start at conjunction at the
companion's aphelion, their closest approach distance attains its
largest possible value.  One would therefore expect such an initial
condition to lead to an increase in the planet's survival time.  In
contrast, if the planet and companion start at conjunction at the
companion's perihelion, their closest approach distance attains its
smallest possible value.  This initial condition therefore has a
destabilizing effect on the planet's orbital motion.  The goal of this
section is to explore what effect the presence of mean motion orbital
resonance conditions has on the survival time distribution for
randomly chosen initial orbital phases, which in turn affects the
structure of output variables generated in this type of analysis.

Toward this end, we first consider a series of low inclination runs
with $M_C = 0.1 \msol$ and $e = 0.5$, for which the ratio of initial
orbital periods is either of the form $n_C:1$ or $n_C:2$, ranging from
$7:1$ to $14:1$ (series 5).  The results of these experiments are
presented in Figure 8, which plots the mean value of $\mu_s$ for a
given orbital configuration as a function of periastron.  Values from
$n_C:1$ orbital configurations are represented with solid point-type,
and the value of $n_C$ is marked below the corresponding data.  Values
from $n_C : 2$ orbital configurations are represented with open
point-type.  As expected, the mean $\langle\mu_s\rangle$ exhibits an
overall increase with increasing periastron, but for orbital
configurations with $\langle\mu_s\rangle \simgreat 4$, the survival
times for cases with $n_C:1$ ratios are clearly shorter than those of
their $n_C:2$ counterparts.  Interestingly, both cases correspond to
necessary (but not sufficient) condition for resonance (e.g., Murray
\& Dermott 2000).  Our results thus indicate that stabilization is
more likely to occur for integer ratios of the form $n_C:2$ compared
to the $n_C:1$ case.  Indeed, of the 4,200 runs performed in series 5
(100 equivalent realizations for each of the 42 orbital configurations
explored), 38 of the 42 experiments that remained bound over the
$\tau_{run} = 10^9$ yr span had $n_C:2$ orbital configurations.
Further evidence of enhanced stabilization in the $n_C:2$ cases is
provided in Figure 9, which shows the $\mu_s$ distributions for the
$11:1$, $23:2$ and $12:1$ orbital configurations ($i = 0^o$) from
series 5 (we note that the last bin for the 23:2 distribution contains
7 numerical experiments which survived out to the maximum integration
time of $10^9$ years).  While all three distributions show peaks with
widths $\sim 0.5$, the $23:2$ distribution (whose orbital
configuration has a periastron that falls between those of the 11:1
and 12:1 cases) has its peak centered at a considerably larger value
of $\mu_s$ than the others, and has a significant, almost flat tail
above the peak.

While a full exploration of the effects of mean motion resonances in
these three body systems is not computationally feasible, we perform a
series of runs (series 6) focusing on the $11:1$ to $12:1$ region of
series 5.  The results of these experiments are presented in Figure
10, with the mean value of the $\mu_s$ distribution for a given
orbital configuration plotted as a function of periastron at $i = 0^o,
10^o$, and $20^o$.  Figure 10 suggests that a significant amount of
structure can be attributed to $n_C:n_p$ type resonances, and that the
survival time is longest for larger values of $n_p$.  This latter
result is further suggested by the plot of the width of the $\mu_s$
distributions as a function of periastron, as shown in Figure 11.
Clearly, the distributions with the greatest widths are associated
with large $n_p$ type orbits.  Nevertheless, Figure 10 suggests that
the dominant features present in the $\langle\mu_s\rangle$ - $p$ curve
are broad ``depressions'' which occur around the $n_C:1$ orbits.
Figure 8 further suggests that these ``depressions'' become more
pronounced with increasing periastron (and hence increasing
$\langle\mu_s\rangle$). This issue is complicated and could be the
subject of a considerably broader investigation.

An important consequence of these results is the implied additional
computational cost in numerically determining the survival times for
long-lived systems.  Specifically, it appears that $\mu_s$
distributions can become significantly wider than $\sim 0.5$ for
orbital configurations with longer survival times, e.g., as shown in
Figure 12.  This result thus presents an added challenge to the
numerical exploration of long-lived planetary systems. In addition to
the increased computational time required to perform a given numerical
experiment, the number of equivalent runs that must be performed to
acquire a valid statistical result for each orbital configuration also
increases.

\subsection{Ejection time versus periastron and eccentricity}

Previous work on the dynamical stability of Earth-like planetary
orbits in binary systems (e.g., D03) suggests that the most important
variables affecting the system's survival time are the companion mass
and periastron distance. In the work of D03, two surveys of the $a-e$
plane were performed using two different numerical methods (both B-S
and a symplectic integration scheme). Although the survey of parameter
space was not as extensive at that of the present paper, the results
are in good agreement with those found herein (see below; see also
Holman \& Weigert 1999). The mean survival time of a planet in a
binary system can be described by a function of periastron of the form
\be 
\tau_s = \tau_{s0}\, \hbox{\rm exp} \left[\alpha (p-1)\right]\;,
\label{eq:pfunction} 
\ee
where periastron $p$ is in AU and the values of $\alpha$ and
$\tau_{s0}$ depend on the companion mass (e.g., see bottom panels of
Figures 3 -- 6 and Table 1 in D03). For a given periastron value, the
survival time has a relatively wide distribution and equation 
(\ref{eq:pfunction}) provides a fit to the mean values.  

We note that the width of the distribution of survival times (at
constant periastron) arises from two sources: (1) The chaotic nature
of the system, which leads to an intrinsic width for any unstable
orbital configuration, i.e., for given values of $(a, e)$; and (2) The
sampling over ($a, e$) pairs at constant periastron $p$, which
provides an additional systematic width to the distribution. As
discussed in \S 3, the intrinsic width typically has a value $\sigma
\sim 0.5$ (for the distribution of $\mu_s = \log \tau_s$), although
the tails at long survival times can make the effective total width
larger. The range of survival times, for a given periastron, is wider
still, where the additional variation is due to the systematic width
defined above.

Figure 13 presents an analogous plot of survival time as a function of
periastron $p$ for the simulations of this paper. Specifically, we
plot the mean values of $\mu_s$ versus periastron for our numerical
simulations of the low inclination ($i \le 20^o$) configurations with
$M_C = 0.5 \msun$ (series 2 -- 4). The figure also shows the value of
$\tau_s$ calculated from equation (\ref{eq:pfunction}) using the
values of $\alpha = 4.7$ and $\tau_{s0} = 0.64$ given in Table 1 of
D03 (appropriate for $M_C = 0.5 \msun$).  The results of this present
work are thus in good agreement with those of D03.


We explore more carefully the structure in our output measures arising
from different eccentricities by performing a series of runs with $M_C
= 0.5 \msun$ and $i = 0^o$, sampling over eccentricity for values of
$p$ ranging between 2.5 and 4.5 AU (series 7).  As with series 5 and
6, $\tau_{run} = 10^9$ yrs and $N = 10^2$.  The results of these
experiments are presented in Figure 14, which shows the mean of the
$\mu_s$ distributions versus eccentricity for different periastron
values.  The shapes of the $\mu_s - e$ curves (for a given periastron
value) indicate the presence of competing effects on the stability of
the orbital systems being explored.  Specifically, for a given
periastron, an increased value of eccentricity leads to a longer
orbital time, and hence a longer time between closest approaches. As a
result, one expects an increase in survival time with increasing
eccentricity -- a clear trend in the high $e$ part of the periastron
tracks shown in Figure 14. However, lower eccentricity orbits also
become more stable in spite of the reduced orbital times.  This result
points to the importance of the tangential component of the impulse
imparted on the planet at closest approach on destabilizing the
system.  As a result, low eccentricity orbits (for which these
tangential components are small) can have very long survival times.
Indeed, a comparison between $e = 0.2$ and $0.8$ orbits (with all
other orbital parameters the same), as shown by the $\mu_s$
distributions in Figure 15, illustrates how dramatic this effect can
be.

As with the widening of distributions due to the presence of integer
ratios of orbital periods, the stabilization of low eccentricity
orbits poses a two-fold challenge for the numerical analysis of such
systems. An increasing survival time requires longer integration
times, and an increasing width of the distribution requires an
increase in the number of equivalent experiments that must be
performed in order to gain good statistical output measures.  On the
other hand, binary systems are likely to have large eccentricities. 
Specifically, the eccentricity distribution for binaries with orbital
periods $P_b \ge $ 1000 days (e.g., $a_C \ge 2$ AU) has the form $f(e)
= 2e$ (Duquennoy \& Mayor 1991), so that low eccentricity orbits are
considerably less common than their high-eccentricity counterparts.

\subsection{Ejection versus accretion events}

In this section we consider the presence of structure in the output
measure $f_e$, the fraction of events that are ejected from the system
(excluding events that remain bound for the duration of the numerical
run-time $\tau_{run}$).  In order for ejection to occur, the
gravitational energy between the planet and companion at closest
approach must exceed the corresponding gravitational energy between
planet and Sun.  Since the nearest possible approach $d$ between
planet and companion is $d \sim p - 2$ AU (when the planet's
eccentricity is near unity), we find that ejection is not
energetically possible when $p \simgreat 2 + 2M_C/M_\sun$, or $p =
2.2$ AU for series 1, 5, and 6 and $p = 3$ AU for series 2 -- 4 and
7. This result agrees relatively well with the results of our
numerical experiments, especially for low inclination angles.  Our
results also indicate a clear dependence for $f_e$ on inclination
angle, as can be seen from Figure 16, which plots $f_e$ versus
inclination angle at different periastron values for series 1 -- 4.
As a general trend, ejection fraction values peak at inclination
angles ranging between $30^o$ and $60^o$, but additional structure is
seen in the $f_e - i$ curves.

\subsection{Kozai resonance}

Figure 1, panel (d) shows a clear peak at $\sim 60^o$, a feature that
can be attributed to the presence of a Kozai resonance (e.g., Murray
\& Dermott 2000).  This type of resonance exists for low-mass objects
in highly inclined orbits perturbed by an larger mass object with an
outer, nearly circular orbit (which explains why the feature is not
observed at higher companion eccentricities).  At a Kozai resonance,
the eccentricity and inclination angle of the planet are coupled such
that when one is at its maximum value, the other is at its minimum
(and vice versa), and the quantity
\be
H_K = \left(a_C\,\left[1-e^2\right]\right)^{1/2} \; \hbox{\rm cos}\; i
\ee
remains constant (Kozai 1962).  

To confirm the presence of Kozai resonances in the $\sim 60^o$ runs of
series 4, we numerically determine both the planet's eccentricity and
inclination angle as a function of time for three different test runs,
defined by the orbital parameters: Run 1 -- $M_C = 0.5 \msun$, $p = 4$
AU, $e = 0.75$ and $i = 0^o$; Run 2 -- $M_C = 0.5 \msun$, $p = 4$ AU,
$e = 0.75$ and $i = 60^o$; Run 3 -- $M_C = 0.5 \msun$, $p = 4$ AU, $e
= 0.25$ and $i = 60^o$.  The first two runs have orbital
configurations that should not lead to the presence of Kozai
resonances owing to a small inclination angle and/or a large
eccentricity, whereas Run 3 meets the required conditions for Kozai
resonances to occur.  The results of the three simulations are shown
in Figure 17, which plots the eccentricity for the planet as a
function of time (over the planet's survival time) in panels (a) --
(c), and the eccentricity (solid line) and inclination angle (dotted
line) for a short time interval of Run 3 in panel (d).  The latter
panel shows a clear periodicity in eccentricity and inclination angle
as expected during a Kozai resonance.  We also plot the value of $H_K$
as a function of time for each run in Figure 18.  The value of $H_K$
clearly oscillates between 0.23 and 0.27 for most of the duration of
Run 3, indicating that while the system is not perfectly in resonance
(since $H_K$ is not exactly constant), it is close to it.  In
contrast, the values of $H_K$ for Runs 1 and 2 show considerably more
variability.

\section{CONCLUSION}  

This paper extends previous work on numerical simulations of
Earth-like planets in binary systems.  Specifically, the survival time
of an Earth-like planet orbiting a Sun-like star in the presence of an
outer companion is determined for different orbital configurations,
defined by the companion's mass, eccentricity, periastron, and the
planet's inclination angle relative to the binary orbital plane.  Due
to the chaotic nature of the systems being explored, the survival time
for each orbital configuration was determined for $N$ equivalent
realizations with randomly selected initial phase angles, allowing for
a statistical analysis of the survival time.  We used $N = 10^3$ for
the bulk of our exploration of parameter space, which focuses on four
pairs of companion masses and eccentricities (series 1 -- 4), and $N =
10^2$ for three additional series of runs (5 -- 7).  In all, we
performed $\sim 400,000$ numerical integrations and explored $\sim
500$ different orbital configurations.  Our results indicate that the
values of survival time obtained for a given orbital configuration are
log-normally distributed (to leading order).  We therefore use the
logarithm of the survival time $\mu_s \equiv $ log $\tau_s$ (where
$\tau_s$ is given in years) as our primary output measure. 

We plot the dependence of $\langle\mu_s\rangle$ on inclination angle
and periastron value for each distribution in series 1 -- 4. These
results confirm a weak dependence of survival time on inclination
angle for cases with $i \simless 40^o$, and the well known decrease in
survival time at high ($i \simgreat 40^o$) inclination angles.  A
simple extrapolation of our results for low inclination angles
indicates that while potentially Hill unstable, orbital configurations
with a companion mass of $M_C = 0.1 \msun$ and $M_C = 0.5 \msun$ can
nonetheless remain stable for $\sim 5$ Gyrs when $p \simgreat 4$ AU
and $p \simgreat 6$ AU, respectively.  Note that these estimates are
conservative in that the extrapolation is carried out along the lower
envelope of the (wide) range of survival times for a given periastron.
These results are consistent with those of D03, and represent lower
periastron values than those obtained through a Hill stability
analysis (e.g., Gladman 1993; see also Barnes \& Greenberg 2006),
i.e., systems that are ultimately ``unstable'' according to analytic
criteria can survive over the 4.6 Gyr age of the Solar System.

This work has important implications for searches for Earth-like
planets (e.g., Terrestrial Planet Finder) in extrasolar systems, since
a large fraction of stars are found in binaries. As summarized above,
our numerical results show that all binaries with periastron greater
than 6--7 AU allow Earth-like planets to be stable over the age of our
Solar System. This finding, in conjunction with the observed
distributions of binary parameters for solar-type primaries (see
Duquennoy \& Mayor 1991 and the compilation of D03), indicates that
roughly half of all binaries are wide enough to contain stable Earths.
In addition, binaries that are wide enough (with sufficiently large
periaston) to allow for orbital stability of Earth-like planets are
also wide enough to allow for the formation of terrestrial planets
through the accumulation of planetesimals (e.g., Quintana et al. 2002;
Quintana 2004; Marzari \& Scholl 2000). Thus, at least half of the
binary systems are habitable according to dynamical considerations.

Some of the extrasolar planets detected to date are found in binary
systems, although these planets have masses comparable to that of
Jupiter.  In the simulations performed here, the Earth-like planet
acts like a test particle, but even a Jovian mass planet will act as a
test particle in the potential of stellar bodies. As a result, one can
directly apply our stability criteria to these systems: In order for a
Jovian planet to remain stable over typical stellar ages of $\sim5$
Gyr, the binary periastron $p_B$ must be wider than about $\sim7$
times the semimajor axis $a_J$ of the Jovian planet, i.e., $p_B
\simgreat 7 a_J$.  In all of the binary systems with detected planets,
the periastron satisfies this constraint.  A related question is
whether or not an Earth-like planet can survive in a binary system
that also contains a Jovian planet. Although the relevant 4-body
simulations must be left for future work, a rough criterion can be
formulated using our results to date: The Jovian planet must have a
sufficently large periastron to allow the Eath-like planet to remain
stable (and this constriant is roughly $p_J \simgreat 2.5 - 3$ AU,
e.g., D03) and the binary periastron $p_B$ must be large enough to
allow the Jovian planet to be stable (roughly $p_B \simgreat 7 p_J$).

The next major result of this work concerns the width of the
distributions of survival time. We have calculated the $\mu_s$
distributions over a broad range of orbital parameter space. It is
well known that orbital systems can exhibit chaotic behavior in or
near the unstable regime.  As a result, it is not possible to define a
unique value of survival time for a given orbital configuration.
Instead, the survival time can only be defined in terms of a
distribution -- the mean and width of which then characterizes the
underlying dynamics of the system.  For the orbital configurations
explored in series 1 -- 4, the resulting $\mu_s$ distributions exhibit
Gaussian peaks with widths $\sigma \sim 0.5$ and generally display
high-end tails when $\langle\mu_s\rangle \,\simgreat 4$.  In this
regime, the value of $\langle\mu_s\rangle$ calculated via $N$
equivalent realizations of a given orbital configuration has a 68\%
probability of being within $0.05 \,\sqrt{(100/N)}$ of the true value,
and a 99\% probability of being within $0.15 \, \sqrt{(100/N)}$ of the
true value.  The situation is more complex for orbital configurations
whose $\mu_s$ distributions have means in excess of $\sim 4$.  While
exhibiting peaks with widths $\sim 0.5$, these distributions may also
have significant tails, thereby leading to larger calculated total
width values ($\sim 0.7 - 2$).  Part of this increased complexity can
be traced to the increasing importance of mean motion orbital
resonance effects.  Additionally, the loss of a tangential component
in the impulse imparted as the planet moves through closest approach
for low eccentricity cases can also lead to fairly broad
distributions, although the paucity of observed low eccentricity
binary systems somewhat limits the importance of this latter effect.
Nevertheless, our results indicate that the determination of survival
times for long-lived systems poses a serious numerical challenge in
that both the integration times must be longer and (because of the
broader distributions) a larger number of equivalent realizations must
be performed in order to get good statistical results.

The form of the distributions of survival times is not unexpected.
Whenever a large number of independent variables play a role in
determining a distribution, the result tends to exhibit a nearly
gaussian form. In the limit of an infinite number of variables, the
central limit theorem indicates that the resulting distribution will
be gaussian, for any type of distribution for the input variables
(e.g., Richtmyer 1978). Binary systems with planets contain an
intermediate number of variables. Although the systems are sufficiently
complicated to display interesting and complex behavior (Figs. 1 --
18), the number of independent variables in not infinite. As a result,
one expects the distribution of any composite variable (here the
survival time for a given orbital configuration) to approach a
gaussian form, but retain non-gaussian tails (since convergence is
slowest in the tails). The result -- a gaussian distribution with
tails -- is in fact what we find here.

As expected from the chaotic nature of the systems being explored, our
output measures exhibit a rich amount of structure.  Specifically, we
find that the presence of Kozai resonances result in a region of
stability at $i \sim 60^o$ when the companion orbit has a low
eccentricity.  In addition, we find a significant amount of structure
in the output measure $f_e$ - the fraction of events that lead to
ejection for a given orbital configuration.  To our knowledge, this
latter result has not received significant attention in the
literature, and may warrant further study.

\bigskip
\centerline{\bf Acknowledgments} 

We thank Matt Holman for useful discussions.  This work was supported
at Xavier University by the Hauck Foundation, and by the National
Science Foundation under Grant No. 0215836.  This work was supported
at the University of Michigan by the Michigan Center for Theoretical
Physics, and by NASA through the Terrestrial Planet Finder Mission
(NNG04G190G) and the Astrophysics Theory Program (NNG04GK56G0).
Finally, we thank the referee
for a comprehensive set of comments that improved the manuscript.


\newpage
\begin{figure}
\figurenum{1}
{\epsscale{0.90} \plotone{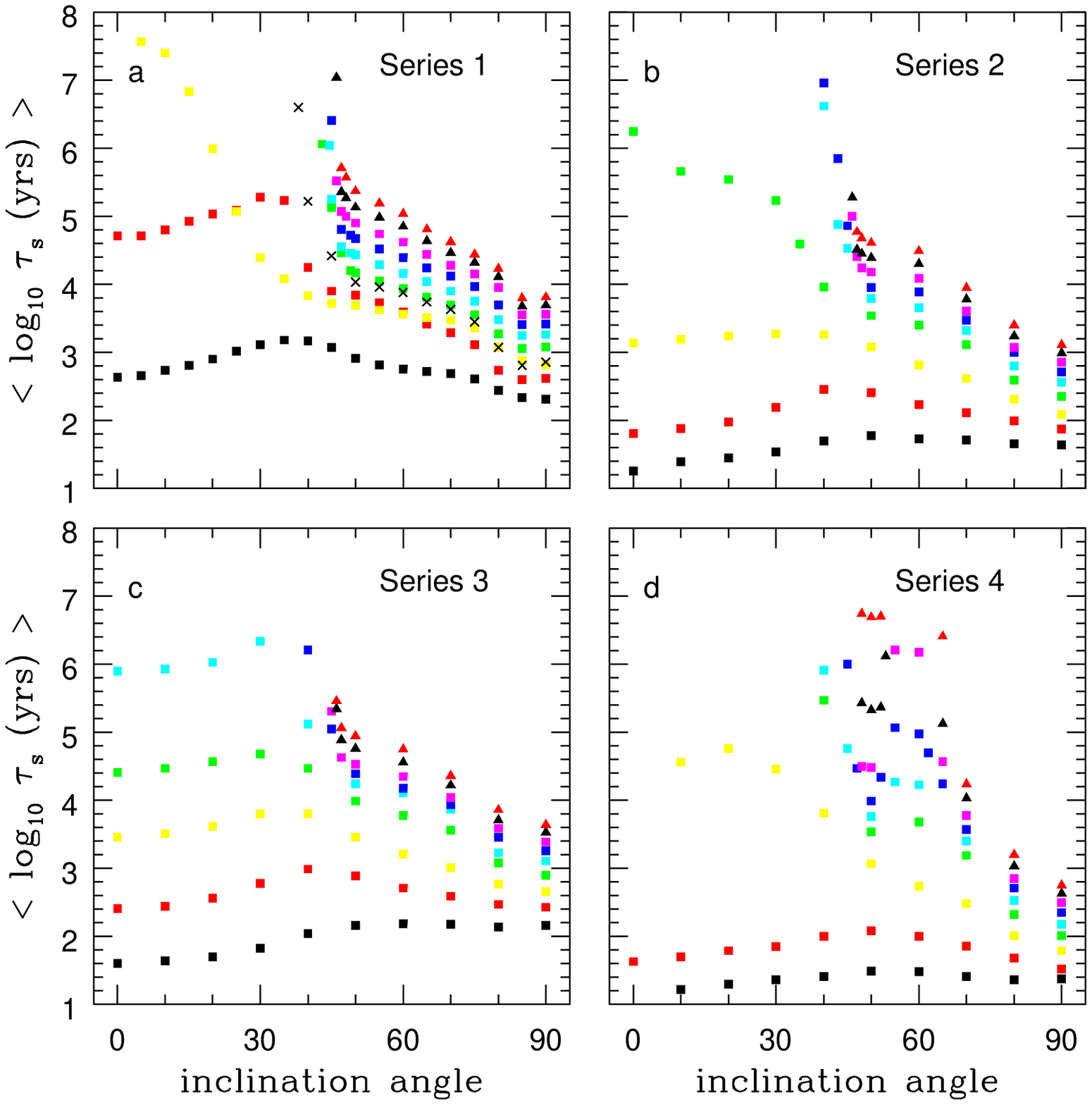} }
\figcaption{Results of numerical simulations for each of the series 1 -- 4, as labeled
in each panel. 
The mean of the $\mu_s$ = log $\tau_s$ distribution generated for each orbital 
configuration is plotted versus inclination angle (degrees) for periastron
values (in AU) of: 2 (black squares); 2.5 (red squares); 2.9 ($\times$);
3 (yellow squares); 3.5 (green squares); 4 (light blue squares); 4.5
(dark blue squares); 5 (purple squares); 5.5 (black triangles); and
6 (red triangles). }  
\end{figure}

\newpage
\begin{figure}
\figurenum{2}
{\epsscale{0.90} \plotone{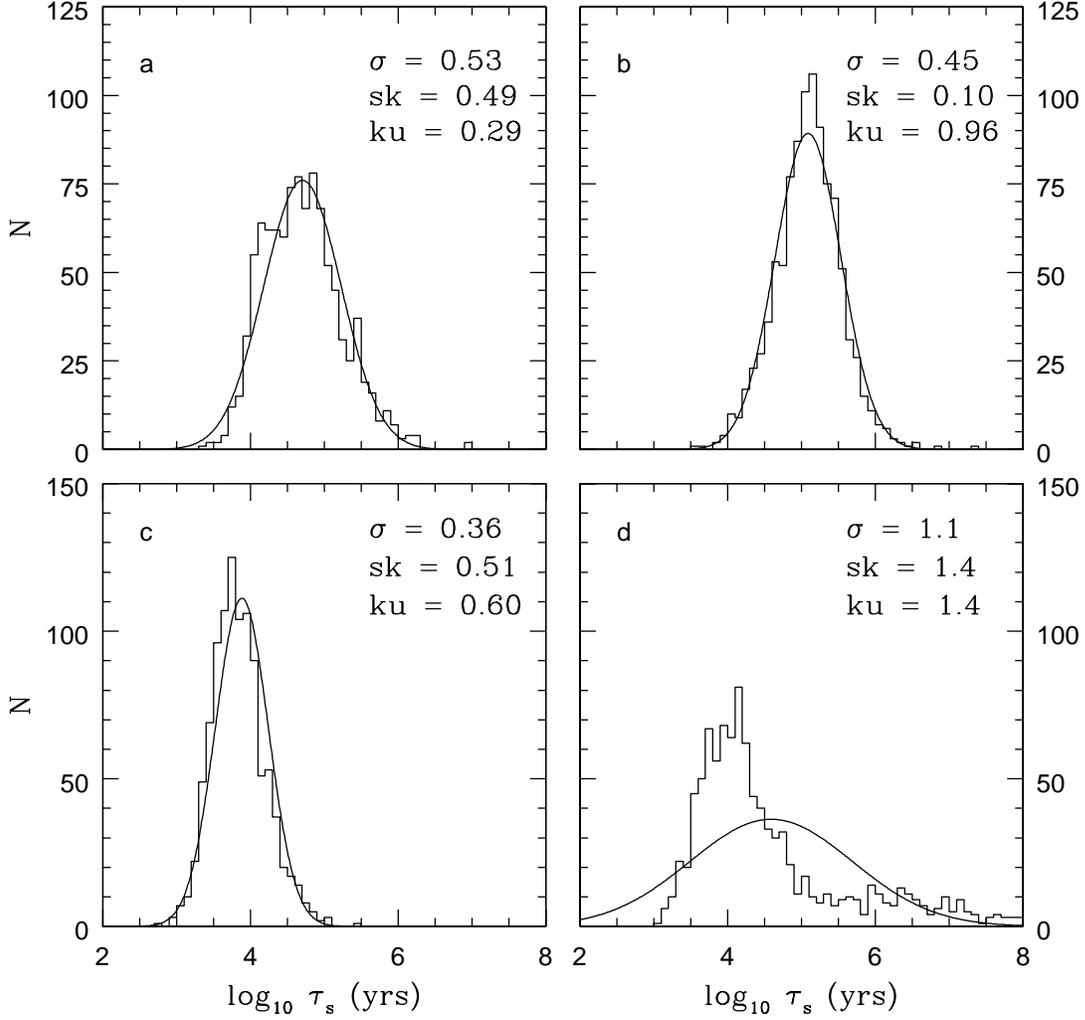} }
\figcaption{The distributions of $N = 10^3$ equivalent values of $\mu_s = $ log $\tau_s$  
(i.e., set through a random sampling of the initial phase angles)
for the following four orbital configurations:
Panel (a) -- $M_C = 0.1 M_\sun$, $p = 2.5$ AU,
$e = 0.5$, and $i = 0^o$; Panel (b) -- same as panel (a), but with $i = 25^o$;
Panel (c) -- same as panel (a), but with $i = 50^o$;  Panel (d) --
$M_C = 0.5 M_\sun$, $p = 3.5$ AU,
$e = 0.5$, and $i = 35^o$. The calculated distribution width, skew  and kurtosis for each
distribution is shown in each panel.  
The solid curve
shows a normal distribution with the same mean and width 
as the computed distributions, normalized to the sample size of 
$N = 10^3$.  
}  
\end{figure}

\newpage 
\begin{figure}
\figurenum{3}
{\epsscale{0.90} \plotone{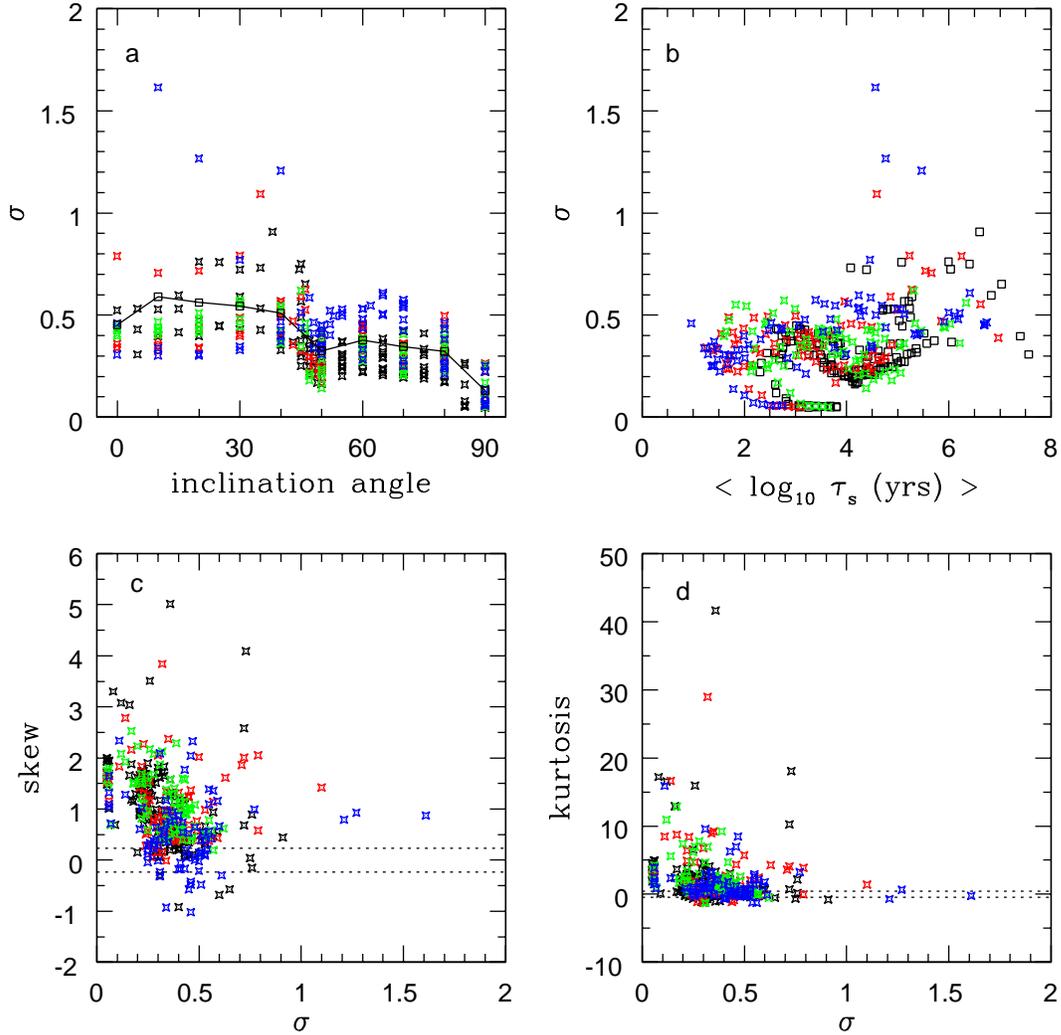} }
\figcaption{Scatter plots of the output measures derived from the 
$\mu_s$ distributions of the
orbital configurations in series 1 -- 4.  Panel (a) -- width versus inclination 
angle; Panel (b) -- width versus mean; 
Panel (c) -- skew versus width; Panel (d) -- kurtosis versus width. 
Results from each series in all four panels are denoted using the 
following color scheme:  series 1 -- black; series 2 -- red; 
series 3 -- green; series 4 -- blue. The solid line in panel (a) 
connects the mean widths calculated at each
inclination angle, determined by combining all of the results from 
orbital configurations with the same inclination angle (but offset by the 
mean so as to center each distribution about zero) to form a single distribution, 
and then calculating the ensuing width as per equation (1). 
The dotted lines in panels (c) and (d) represent
the three sigma values for the skew and kurtosis
distributions, respectively, generated by
randomly sampling a normal parent distribution $N = 10^3$ times.}  
\end{figure}

\newpage 
\begin{figure}
\figurenum{4}
{\epsscale{0.90} \plotone{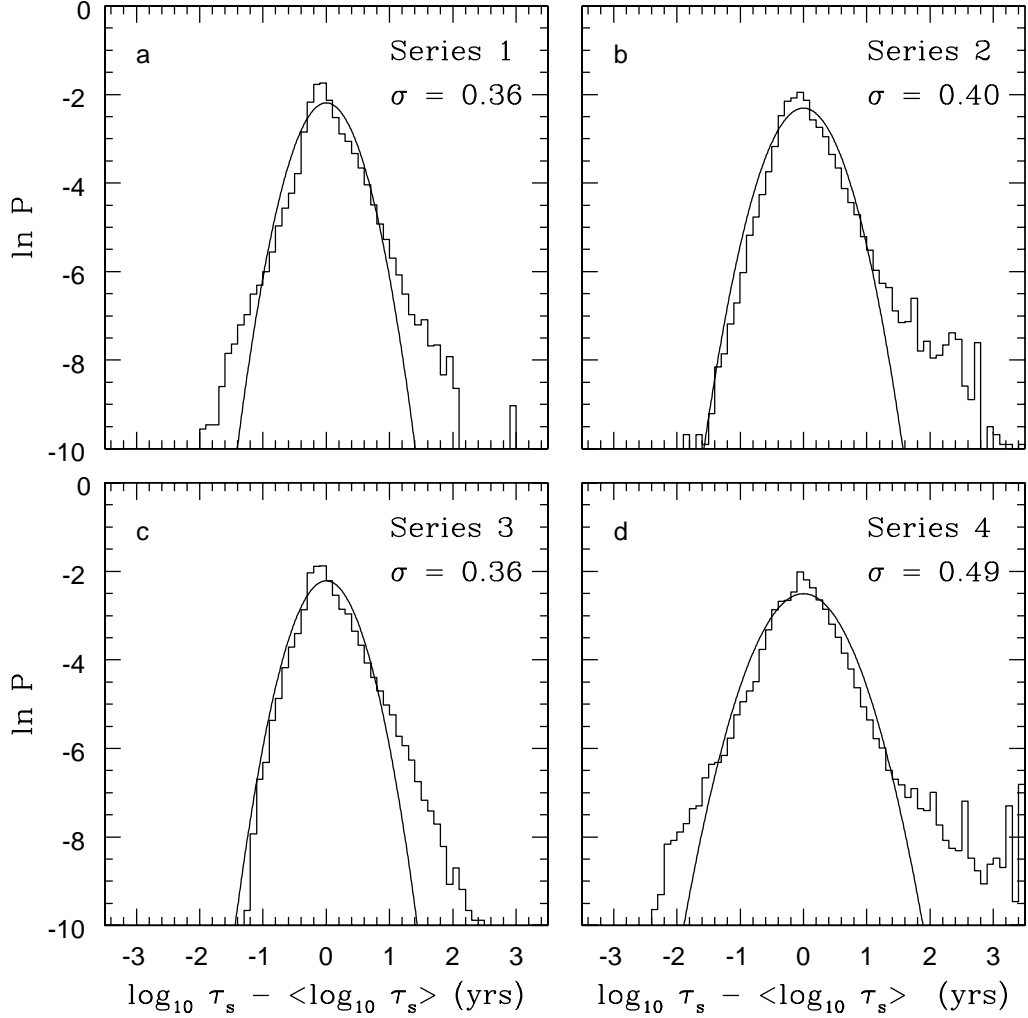} }
\figcaption{The distribution of all of the survival times for orbital 
configurations in: Panel (a) -- series 1; Panel (b) -- series 2;
Panel (c) -- series 3; and Panel (d) -- series 4.  Each value of $\mu_s$
has been offset by the
mean value of its orbital configuration distribution to ensure proper
normalization.
Distributions are plotted in terms of the natural log of the probability
of being in a specified bin. The solid curve in each panel
shows a normal distribution with the same width 
as the computed distributions, normalized to unity.}  
\end{figure}

\newpage 
\begin{figure}
\figurenum{5}
{\epsscale{0.90} \plotone{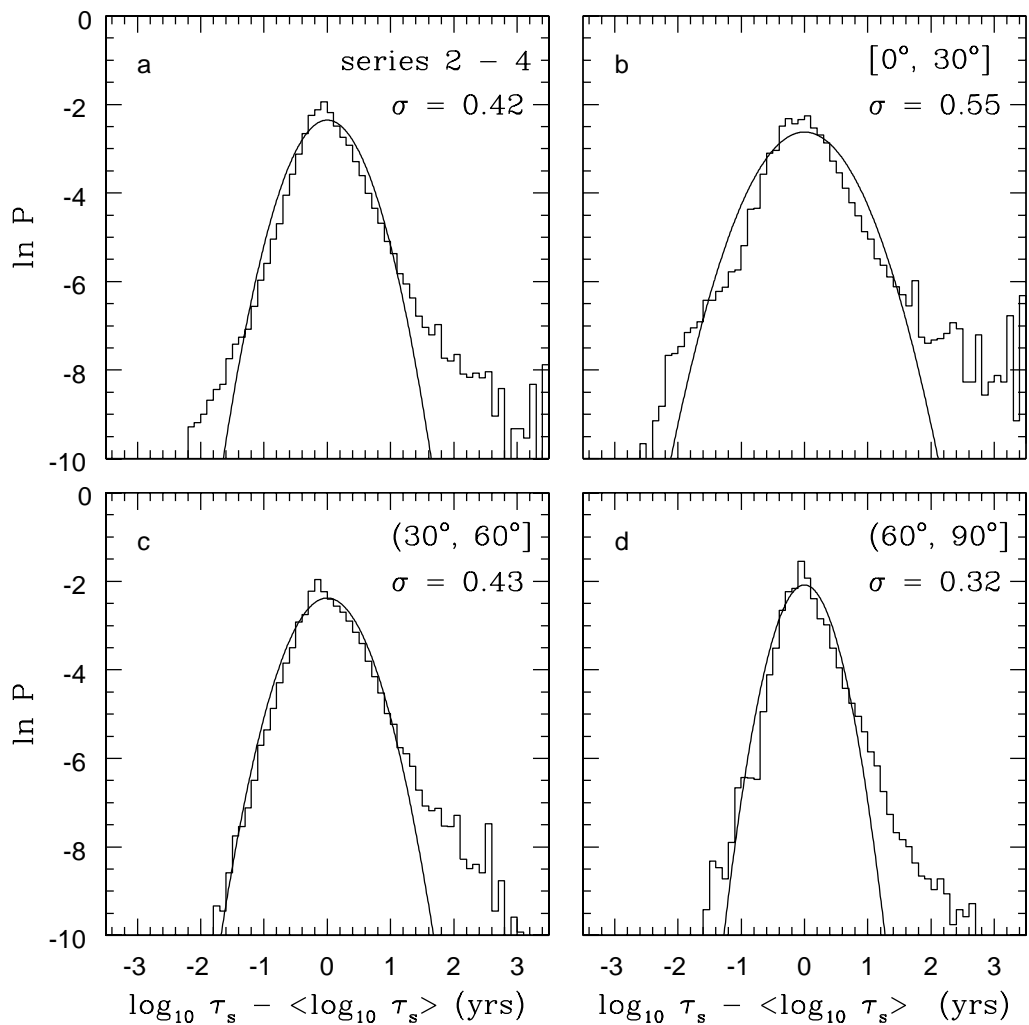} }
\figcaption{Same as Fig. 4, but for: Panel (a) -- all 
orbital configurations in series 2 -- 4; Panel (b) -- 
the subset of series 2 - 4 orbital configurations with
inclination angles $0^o\le i \le 30^o$; Panel (c) -- 
same as panel (b), but for $30^o < i \le 60^o$;
Panel (d) -- same as panel (b), but for $60^o < i \le 90^o$.
}  
\end{figure}

\newpage 
\begin{figure}
\figurenum{6}
{\epsscale{0.90} \plotone{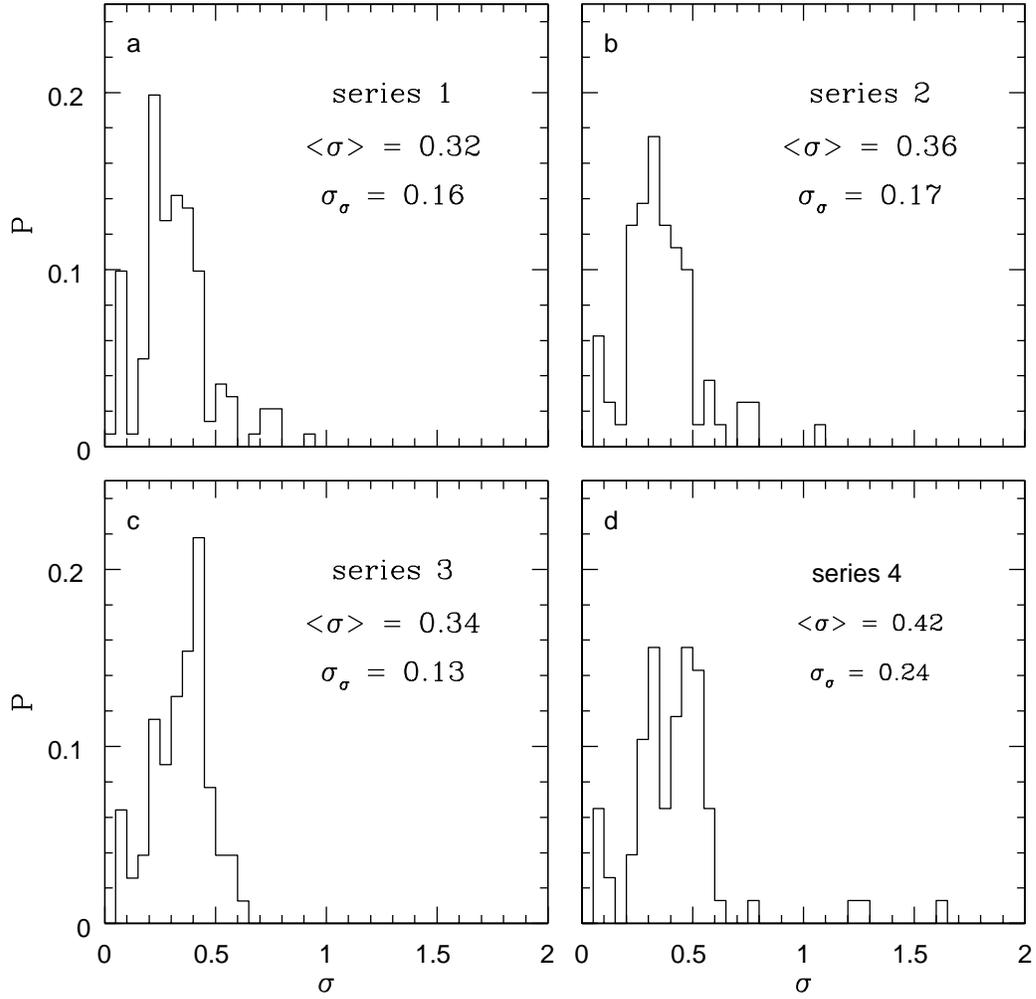} }
\figcaption{The distribution of widths calculated from the
$\mu_s$ distributions for the orbital configurations in: Panel (a) -- series 1; 
Panel (b) -- series 2; 
Panel (c) -- series 3; Panel (d) -- series 4.
The calculated mean $\langle\sigma\rangle$ and width
$\sigma_\sigma$ of the distributions shown in each 
panel are also presented. }  
\end{figure}

\newpage 
\begin{figure}
\figurenum{7}
{\epsscale{0.90} \plotone{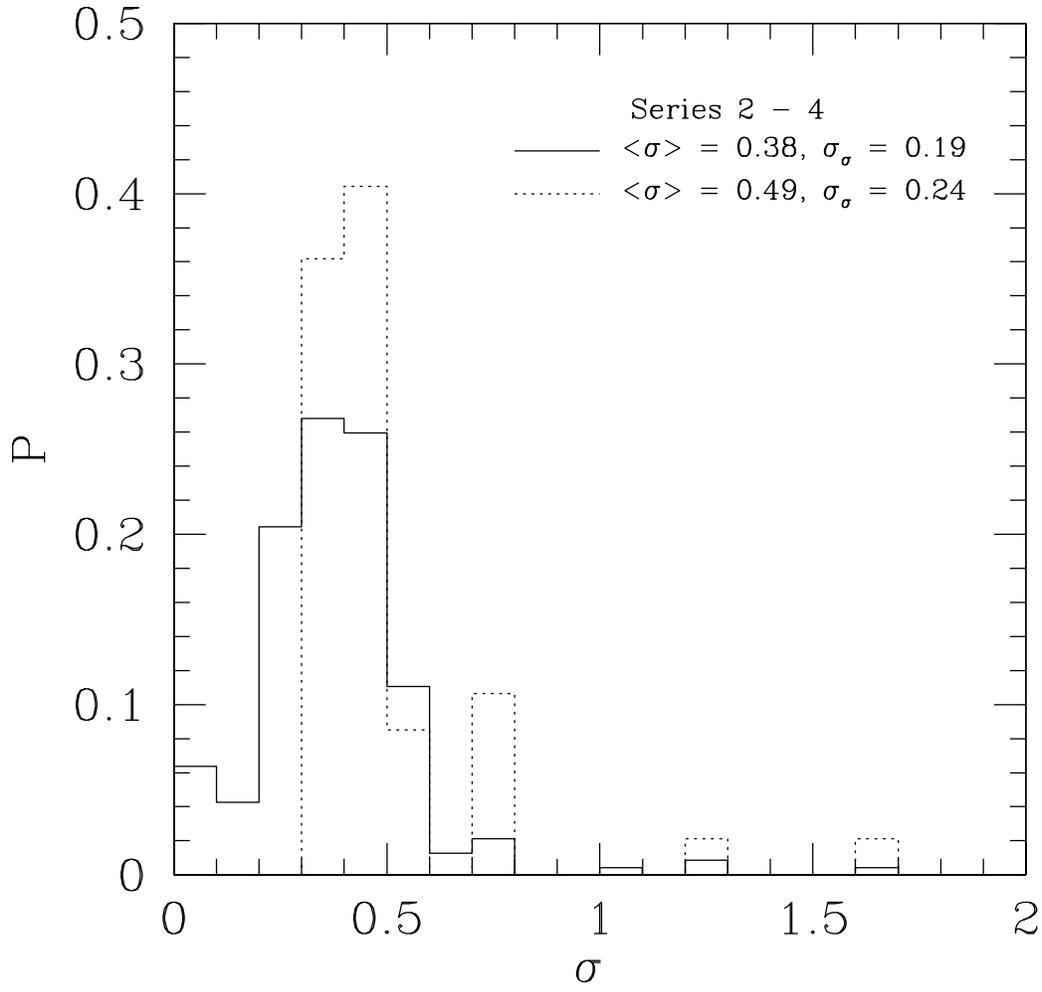} }
\figcaption{Same as Fig. 6, but for all of the orbital configurations
in series 2 -- 4 (solid line) as well as  for the $0^o\le i \le 30^o$ subset 
(dotted line).}  
\end{figure}

\newpage 
\begin{figure}
\figurenum{8}
{\epsscale{0.90} \plotone{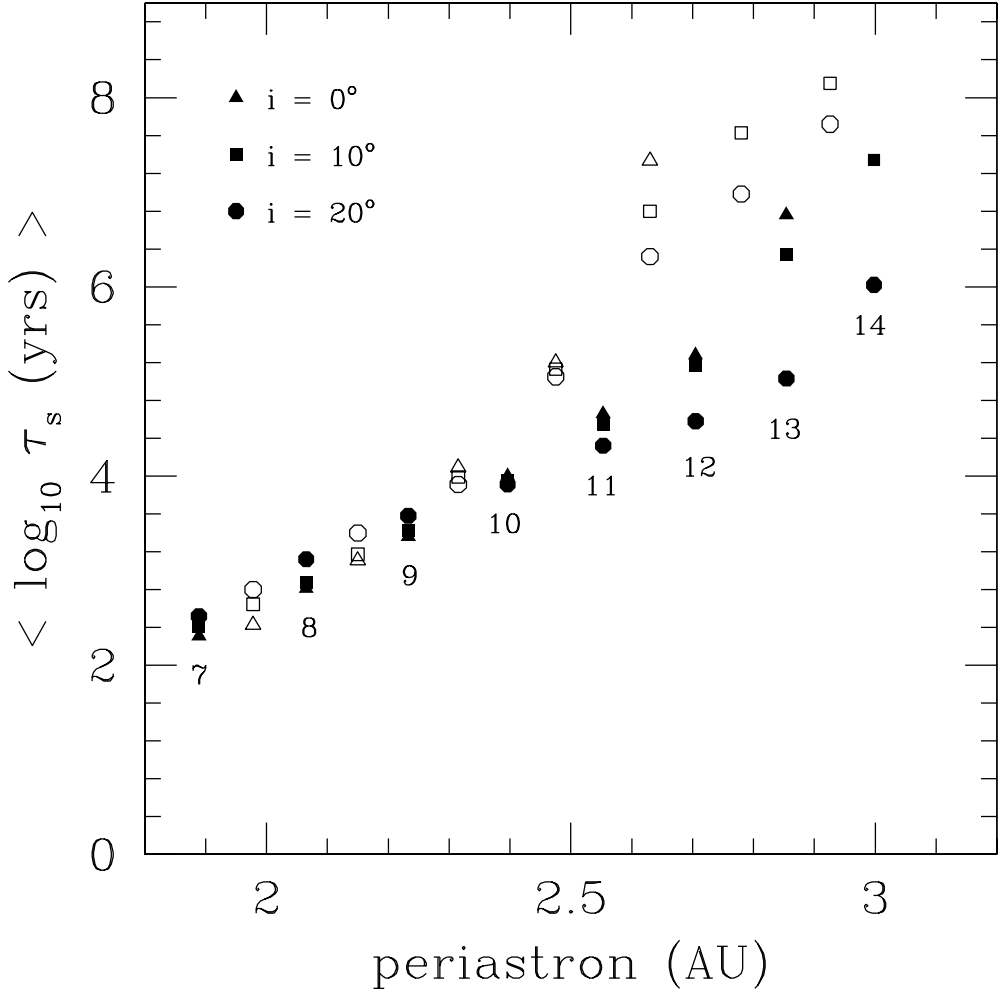} }
\figcaption{Results of numerical simulations for series 5 ($M_C = 0.1 M_\odot$
and $e = 0.5$). The mean of the $\mu_s$ distribution generated for each orbital 
configuration is plotted versus periastron for the following inclination angles:
$0^o$ -- triangles; $10^o$ -- squares; $20^o$ -- circles.  
Values from 
$n_C:1$ orbital configurations are represented with solid points, and the value
of $n_C$ is marked below the corresponding data.  Values from 
$n_C : 2$ orbital configurations are represented by open points, with 
the value of $n_C$ implied by their location.  
}  
\end{figure}

\newpage 
\begin{figure}
\figurenum{9}
{\epsscale{0.90} \plotone{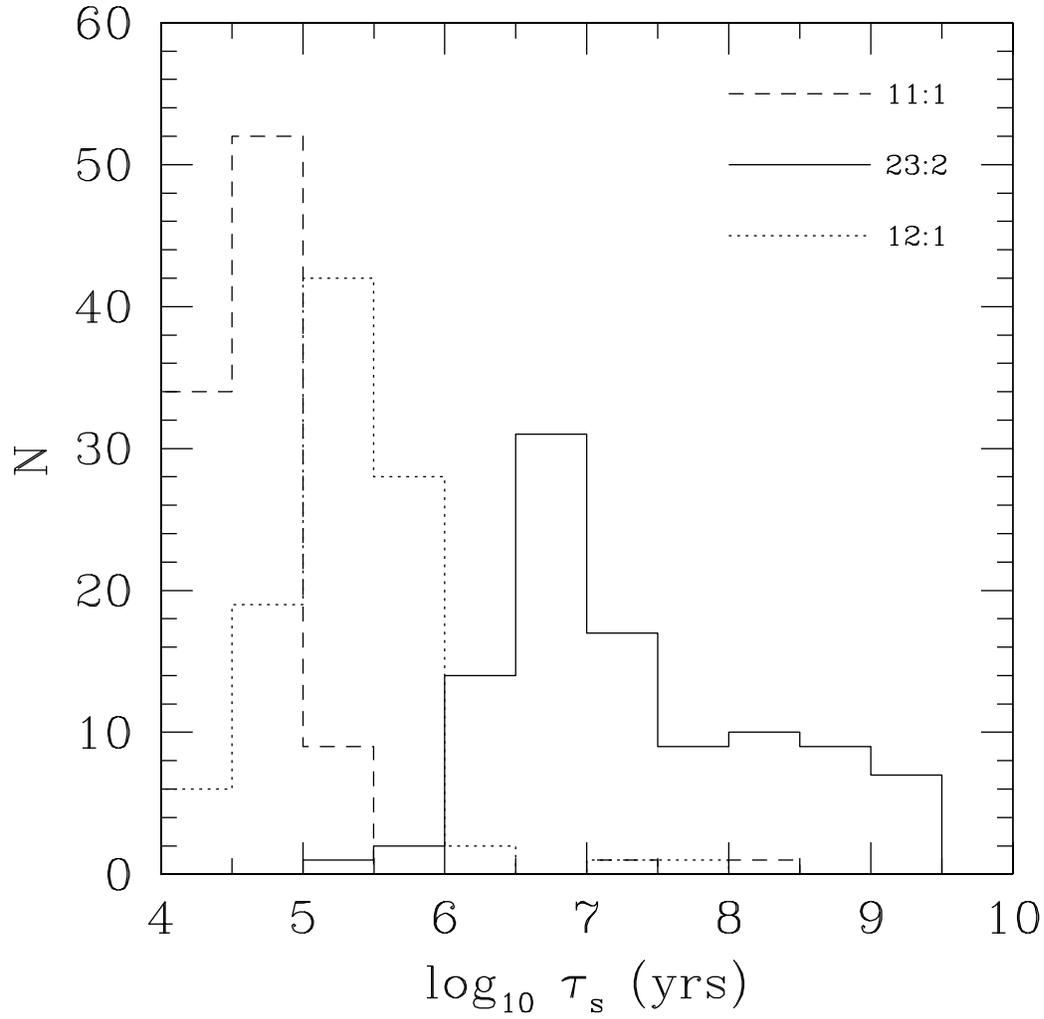} }
\figcaption{The distribution of $\mu_s$ values for three orbital configuration
from series 5.  Long dashed line -- $p = 2.553$ AU ($n_C:n_p = 11:1$);
Solid line -- $p = 2.630$ AU ($n_C:n_p = 23:2$); 
Dotted line -- $p = 2.705$ AU ($n_C:n_p = 12:1$).
}  
\end{figure}

\newpage 
\begin{figure}
\figurenum{10}
{\epsscale{0.90} \plotone{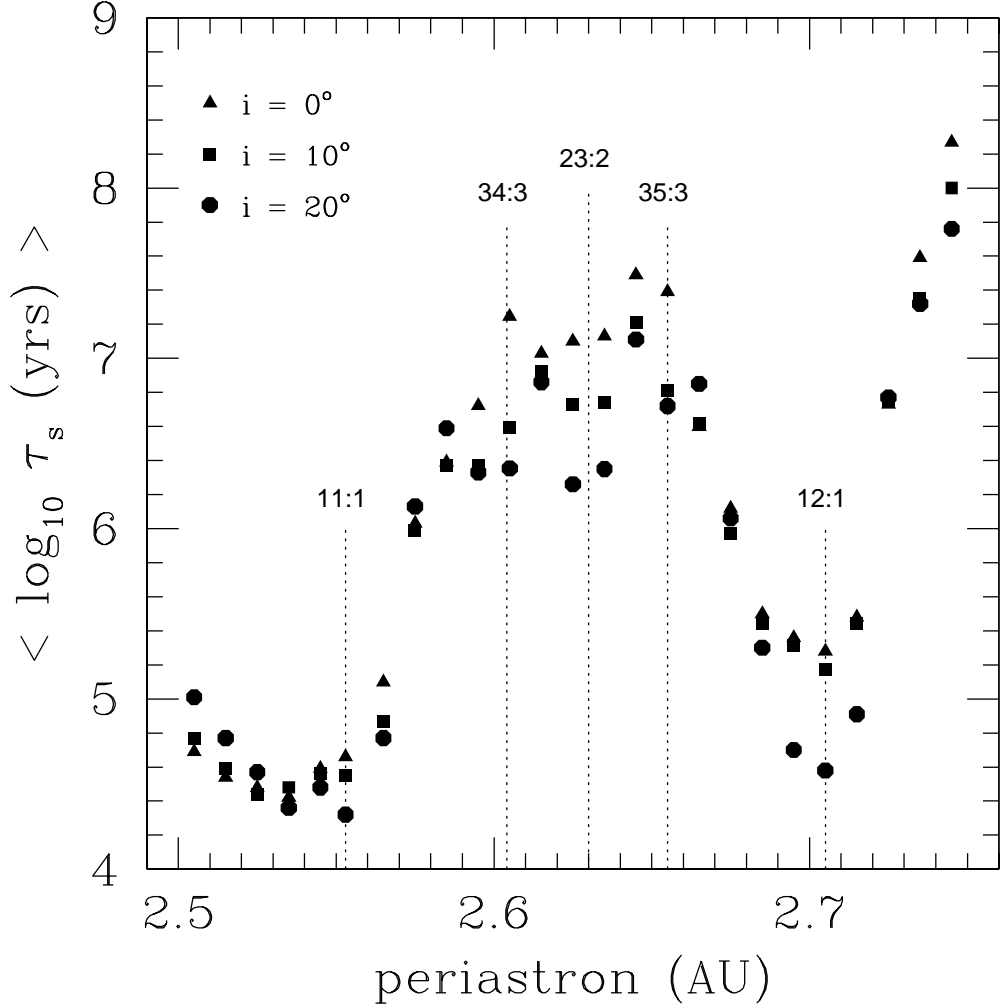} }
\figcaption{Same as Fig. 8, but for a narrower range of periastron values. 
Vertical dotted lines denote periastron values for which integer ratios of 
initial orbital periods occur, with the corresponding $n_C:n_p$ values
labeled above.}  
\end{figure}

\newpage 
\begin{figure}
\figurenum{11}
{\epsscale{0.90} \plotone{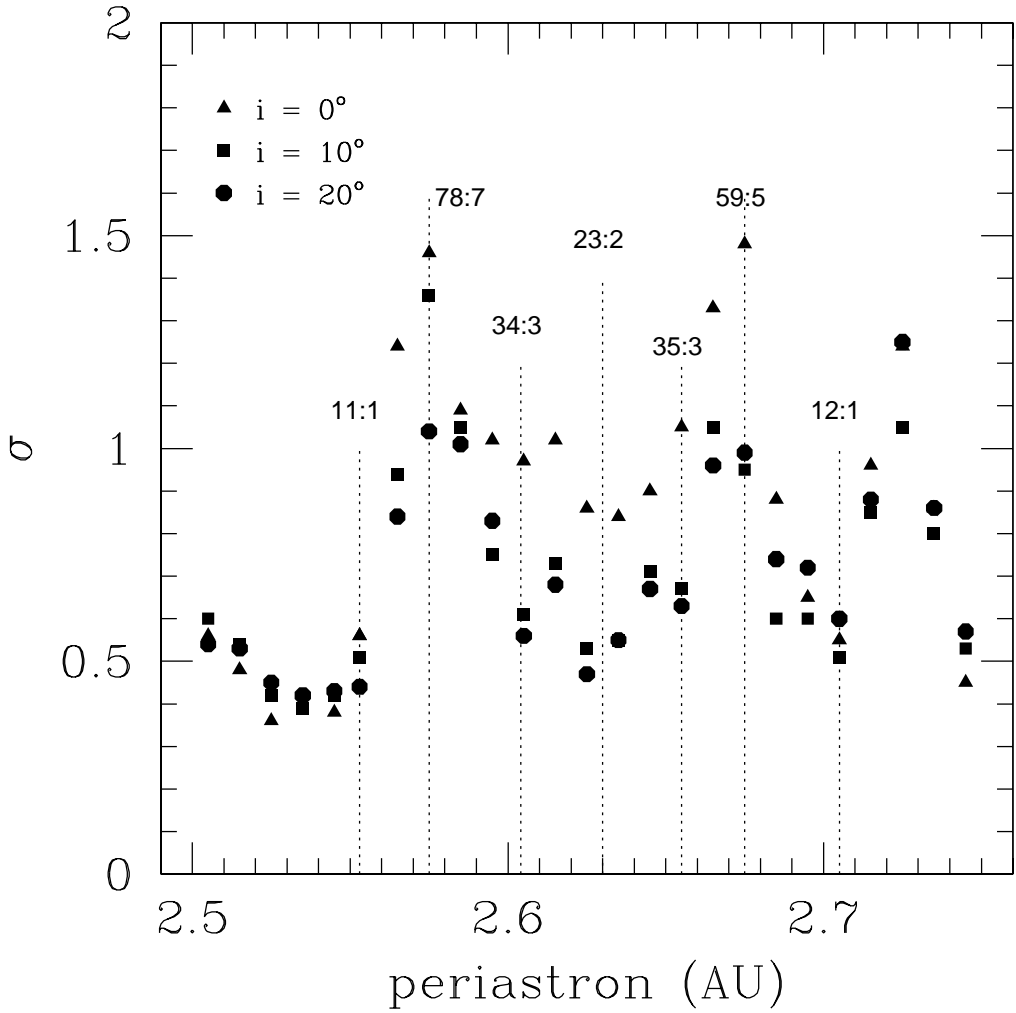} }
\figcaption{The width of the distributions for the orbital 
configurations in series 6
as a function of periastron for the following inclination angles:
$0^o$ -- triangles; $10^o$ -- squares; $20^o$ -- circles.   
Vertical dotted lines denote periastron values for which integer ratios of 
initial orbital periods occur, with the corresponding $n_C:n_p$ values
labeled above. }  
\end{figure}

\newpage 
\begin{figure}
\figurenum{12}
{\epsscale{0.90} \plotone{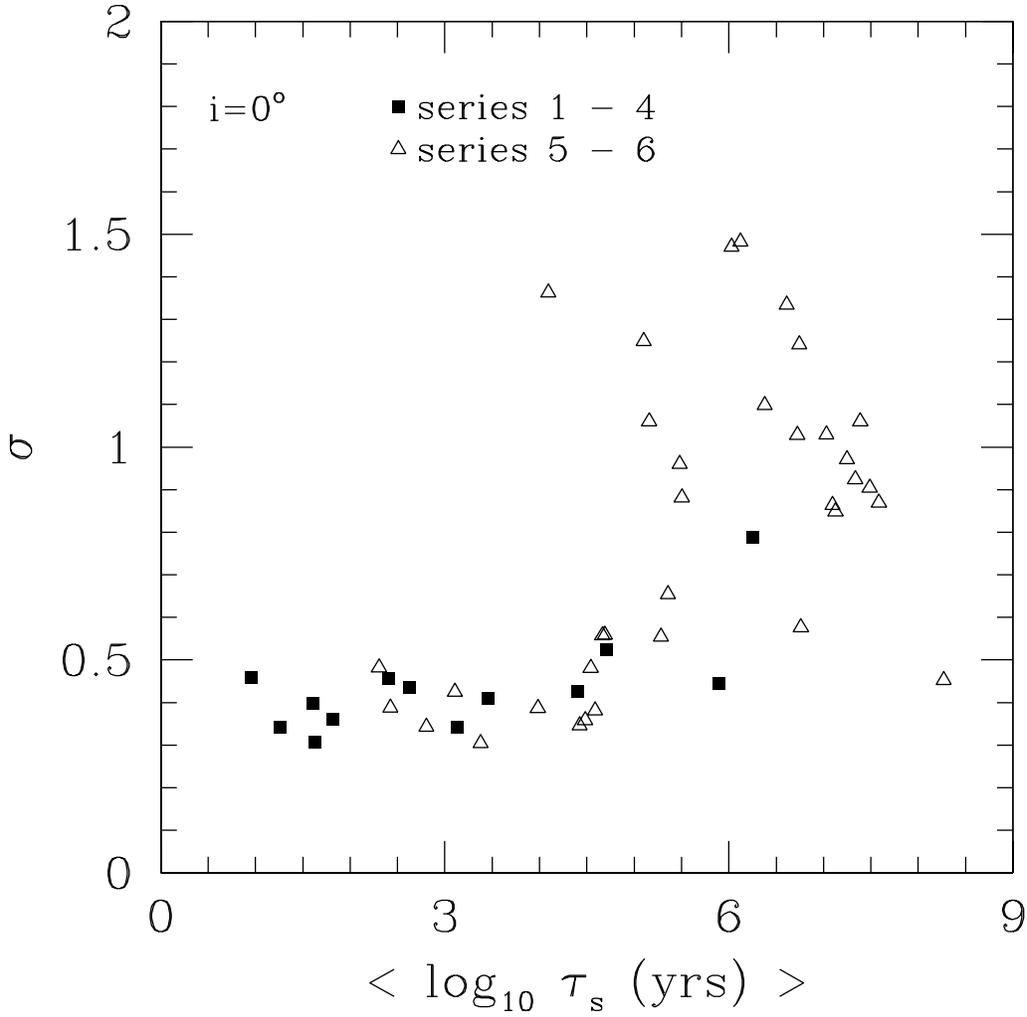} }
\figcaption{The scatter plot of width versus mean for the distributions
of the $i = 0^o$ orbital configurations in series 1 -- 4 (solid squares)
and series 5 -- 6 (open triangles).  }  
\end{figure}

\newpage 
\begin{figure}
\figurenum{13}
{\epsscale{0.90} \plotone{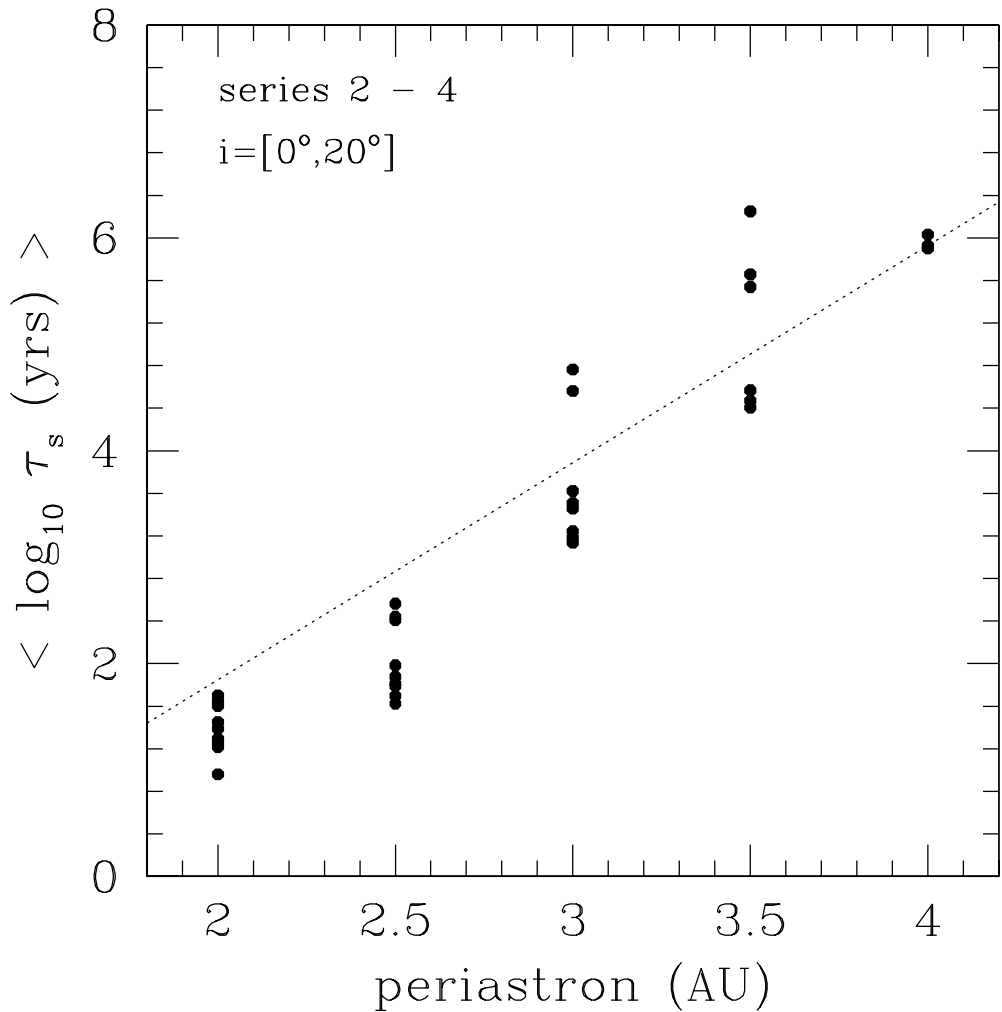} }
\figcaption{Results of numerical simulations for series 2 -- 4 ($M_C = 0.5 M_\odot$). 
The mean of the normal distribution generated for each orbital 
configuration is plotted versus periastron for inclination angles $0^o \le i \le 20^o$. 
The dotted line plots the ejection time given by equation (4) with values of
$\alpha = 4.7$ and $\tau_{s0} = 0.64$ (the best fit values to the corresponding data
presented in D03, Figure 3).}  
\end{figure}

\newpage 
\begin{figure}
\figurenum{14}
{\epsscale{0.90} \plotone{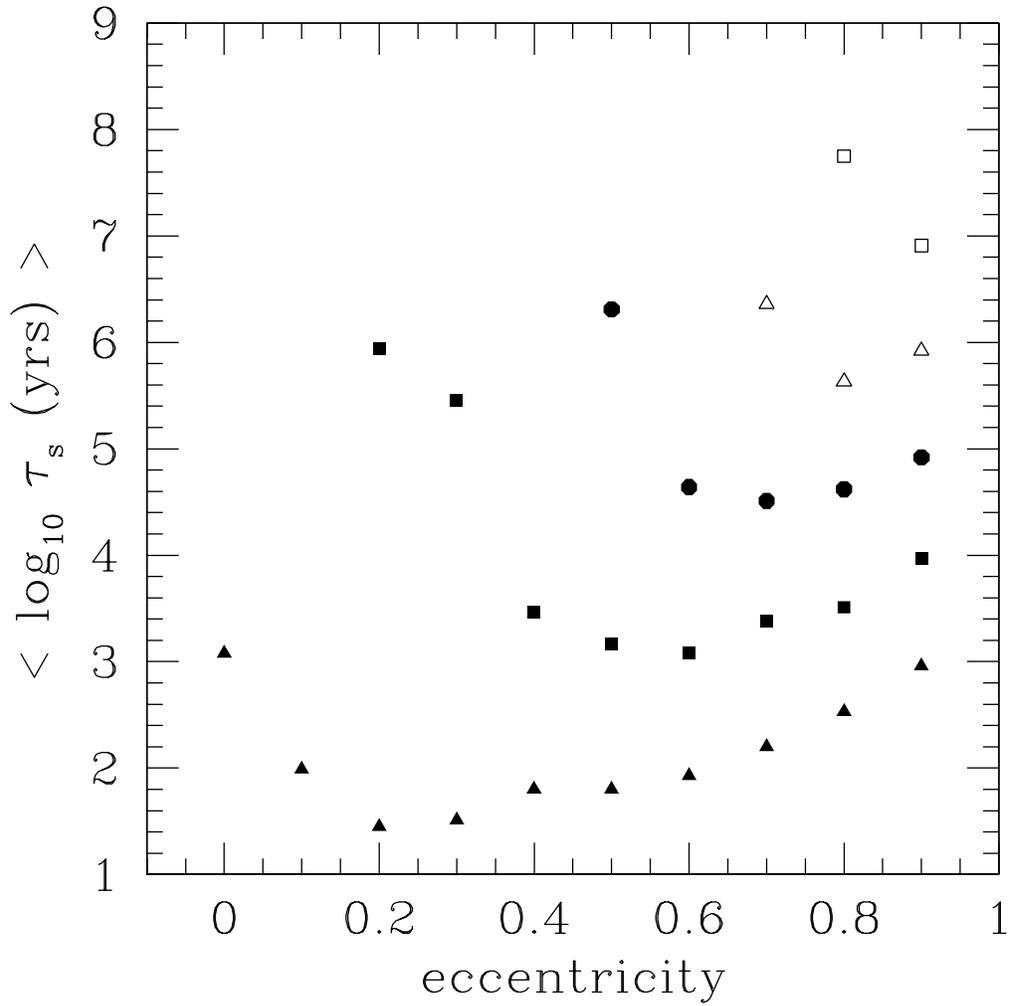} }
\figcaption{Results of numerical simulations for series 7 ($M_C = 0.5 M_\odot$
and $i = 0^o$). The mean of the $\mu_s$ distributions generated for each orbital 
configuration is plotted versus companion eccentricity for periastron values (in AU):
2.5 (solid triangles); 3 (solid squares); 3.5 (solid circles); 4 (open triangles); 
and 4.5 (open squares).}  
\end{figure}

\newpage 
\begin{figure}
\figurenum{15}
{\epsscale{0.90} \plotone{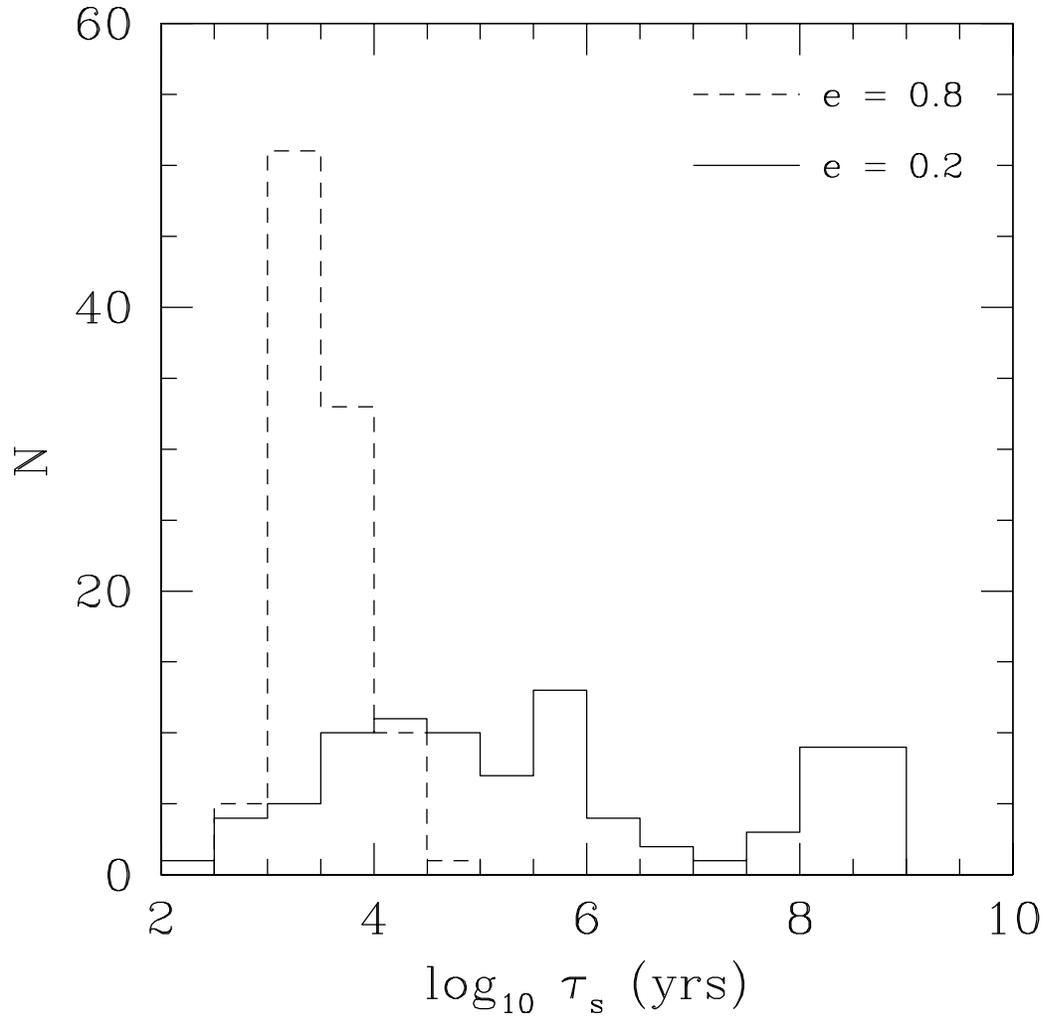} }
\figcaption{The distributions of survival times for two orbital configurations
from series 7, with $M_C = 0.5 \msun$, $p = 3$ AU, $i = 0^o$ and
$e = 0.2$ (solid line) and $e = 0.8$ (dotted line).}  
\end{figure}

\newpage 
\begin{figure}
\figurenum{16}
{\epsscale{0.90} \plotone{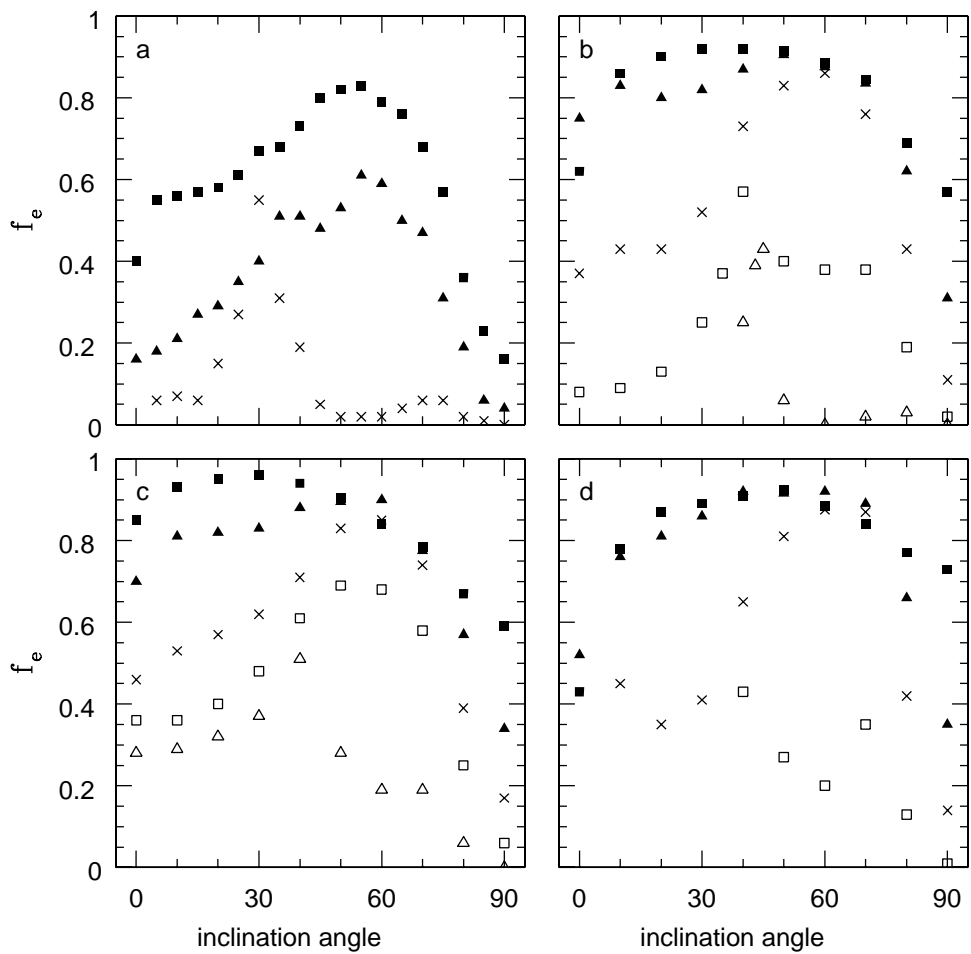} }
\figcaption{The fraction 
of all simulations (excluding those which survive out to the integration time
$\tau_{run}$)
leading to ejection is plotted versus inclination angle for
periastron values (in AU):  2 (solid square); 2.5 (solid triangle); 3 ($\times$);
3.5 (open square); 4 (open triangle).  Panel (a) -- series 1;
Panel (b) -- series 2; Panel (c) -- series 3; and Panel (d) -- series 4.   
}
\end{figure}

\newpage 
\begin{figure}
\figurenum{17}
{\epsscale{0.90} \plotone{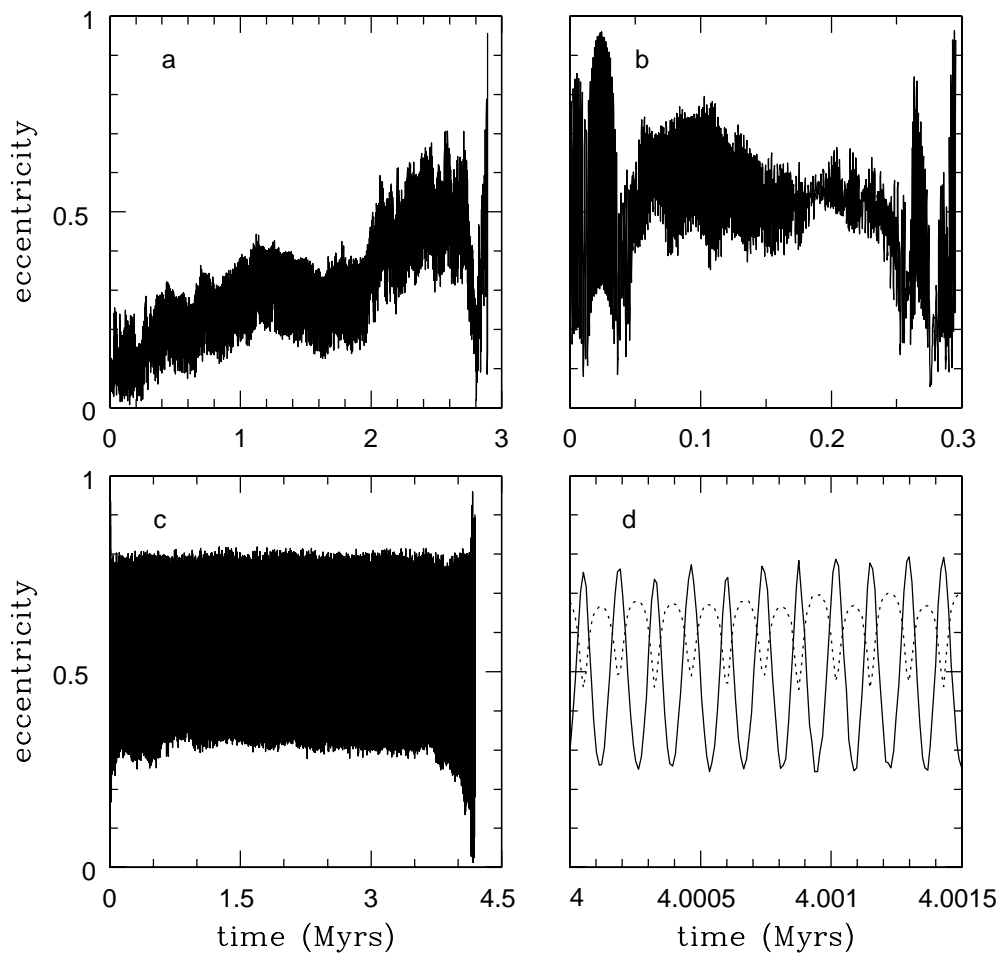} }
\figcaption{The time evolution of the planet's eccentricity
for three numerical experiments
with different orbital configurations. Panel (a) -- Run 1: 
$M_C = 0.5 \msun$, $p = 4$ AU, $e = 0.75$ and $i = 0^o$; 
Panel (b) -- Run 2: $M_C = 0.5 \msun$, $p = 4$ AU, $e = 0.75$ and $i = 60^o$;
Panel (c) --  Run 3: $M_C = 0.5 \msun$, $p = 4$ AU, $e = 0.25$ and $i = 60^o$; 
Panel (d) -- same as panel (c), but for a small time interval of the
planet's evolution.  This panel also shows the planet's
inclination angle (dotted line), as a fraction of $\pi$/2.}  
\end{figure}

\newpage 
\begin{figure}
\figurenum{18}
{\epsscale{0.90} \plotone{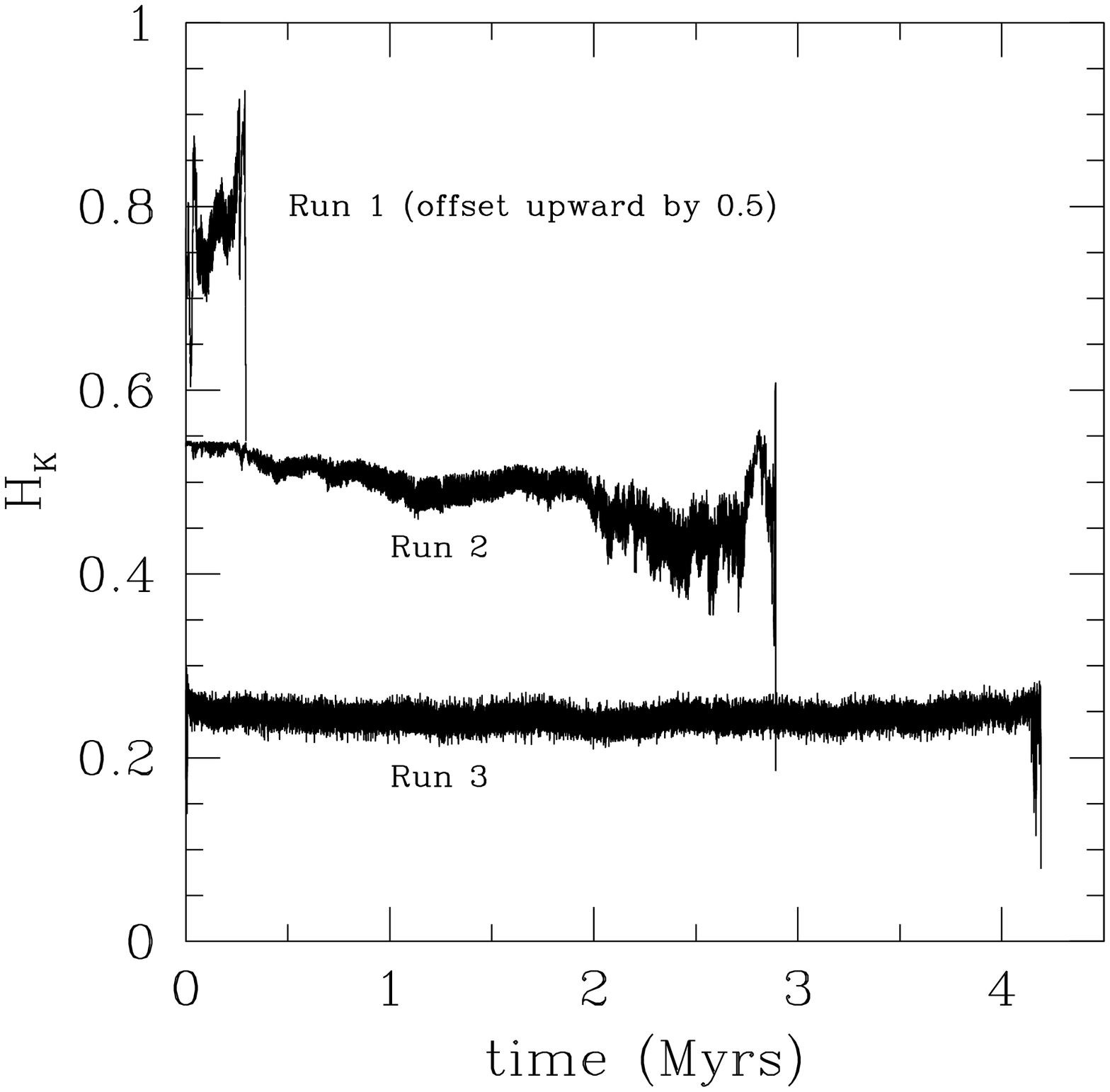} }
\figcaption{The time evolution of $H_K = \sqrt{a_C(1-e^2)}$ cos $i$
for the three numerical experiments presented in Figure 17.  The value 
for Run 1 is offset upward by 0.5.
}  
\end{figure}


\newpage
\begin{thebibliography}{}

\bibitem[as]{as}
Asghari, N., Broeg, C., Carone, L., et al. 2004, A\&A, 426, 353

\bibitem[bg]{bg}
Barnes, R., \& Greenberg, R. 2006, ApJ Lett, in press

\bibitem[b]{b}
Butler, R. P., et al. 1999, ApJ, 526, 916

\bibitem[da]{da}
David, E.-M., Quintana, E. V., Fatuzzo, M., \& Adams, F. C. 2003, 
PASP, 115, 825 (D03) 

\bibitem[du]{du}
Duquennoy, A., \& Mayor, M. 1991, A\&A, 248, 485.

\bibitem[]{}
Dvorak, R., Froeschle, C., \& Froeschle, Ch. 1989, A\&A, 226, 335

\bibitem[eg]{eg}
Eggenberger, A., Udry, S., \& Mayor, M. 2004, A\&A, 417, 353

\bibitem[f]{f}
Fischer, D. A., et al. 2002, ApJ, 564, 1028

\bibitem[g]{g}
Gladman, B. 1993, Icarus, 106, 247 

\bibitem[hag]{hag}
Haghighipour, N. 2006, ApJ, 644, 543

\bibitem[ha]{ha}
Harrington, R. S. 1977, AJ, 82, 753

\bibitem[ho]{ho}
Holman, M. J., \& Wiegert, P. A. 1999, AJ, 117, 621 

\bibitem[in]{in}
Innanen, K. A., Zheng, J. Q., Mikkola, S., \& Valtonen, M. J. 1997, AJ, 113, 1915

\bibitem[jo]{jo}
Jones, B. W., Underwood, D. R., \& Sleep, P. N. 2005, ApJ, 622, 1091

\bibitem[ko]{ko}
Kozai, Y. 1962, AJ, 67, 591.

\bibitem[la99]{la99}
Laughlin, G., \& Adams, F. C. 1999, ApJ, 526, 881


\bibitem[li93]{li93}
Lissauer, J. J. 1993, ARA\&A, 31, 129

\bibitem[li04]{li04}
Lissauer, J. J., Quintana, E. V., Chamnbers, J. E., Duncan, M. J., \& Adams, F. C.
2004, RevMexAA (Series de Conferencias), 22, 99

\bibitem[mar]{mar}
Marcy, G. W., Butler, R. P., Vogt, S. S., et al. 2001, ApJ, 555, 418

\bibitem[]{}
Marzari, F., \& Scholl, H. 2000, ApJ, 543, 328

\bibitem[ma]{ma}
Mathieu, R. D., Adams, F. C., \& Latham, D. W. 1991, AJ, 101, 2184

\bibitem[mc]{mc}
McArthur, B. E., Endl, M., Cochran, W. D., et al. 2004, ApJ, 614, L81

\bibitem[mu]{mu}
Mudryk, L. R., \& Wu, Y. 2006, ApJ, 639, 423

\bibitem[Mur]{MD2000} 
Murray, C. D., \& Dermott, S. F. 2000, Solar System Dynamics
(Cambridge: Cambridge Univ. Press)  

\bibitem[mus]{mus}
Musielak, Z. E., Cuntz, M., Marshall, E. A., \& Stuit, T. D. 2005, 
A\&A 434, 355

\bibitem[no]{no}
Noble, M., Musielak, Z. E., \& Cuntz, M. 2002, ApJ, 572, 1024

\bibitem[pe]{pe}
Pendleton, Y. J., \& Black, D. C. 1983, AJ, 88, 1415

\bibitem[]{}
Pilat-Lohinger, E. \& Dvorak, R. 2002, CeMDA, 82, 143

\bibitem[pi]{pi}
Pilat-Lohinger, E., Funk, B., \& Dvorak, R. 2003, A\&A, 400, 1085

\bibitem[press]{p92}
Press, W. H., et al. 1992, Numerical Recipes: The art of scientific 
computing (Cambridge: Cambridge Univ. Press)   

\bibitem[qu]{qu}
Quintana, E. V., Lissauer, J. J., Chambers, J. E., \& Duncan, M. J. 2002,
ApJ, 576, 982

\bibitem[]{}
Quintana, E. V. 2004, PhD Thesis, University of Michigan

\bibitem[ra]{ra}
Rabl, G., \& Dvorak, R. 1988, A\&A, 191, 385

\bibitem[]{}
Richtmyer, R. D. 1978, {Principles of Advanced Mathematical Physics} 
(New York: Springer-Verlag)

\bibitem[ru]{ru}
Ruden, S. P. 1999, in Origin of Stars and Planetary Systems, ed.
C. J. Lada \& N. D. Kylafis (Dordrech:  Kluwer), 643

\bibitem[su]{su}
S\"uli, \'A, Dvorak, R., \& Florian, F. 2005, MNRAS, 363, 241

\bibitem[]{}
Szebehely, V. 1967, Theory of orbits: The restricted problem of three bodies
(New York: Academic Press)

\bibitem[ti]{ti}
Tinney, C. G., et al. 2002, ApJ, 571, 528 

\bibitem[wi]{wi}
Whitmire, D. P., Matese, J. J., Criswell, L., \& Mikkola, S. 1998,
Icarus, 132, 196

\end{thebibliography}
\end{document}